# L'artefact de la malédiction des ressources naturelles[*]

Matata Ponyo Mapon[†]

Jean-Paul K. Tsasa[‡]

Novembre 2019

**Résumé**
Ce papier réexamine la validité de la malédiction des ressources naturelles, en utilisant la base de données des pays exportateurs de minerais. Les évidences suivantes ont été établies : (i) Les pays riches en ressources naturelles (PRRN) n'affichent pas forcément des performances économiques médiocres; (ii) Les PRRN avec des performances médiocres ont tous un portefeuille d'exportations faiblement diversifié; (iii) En revanche, les PRRN avec un portefeuille d'exportations faiblement diversifié n'affichent pas forcément des performances médiocres. Ensuite, nous développons un modèle d'interactions stratégiques inspiré des jeux bayésiens pour étudier le rôle du leadership et de la gouvernance face aux groupes de pression (« Léviathan ») dans la gestion des ressources naturelles. Nous montrons qu'une amélioration du binôme leadership-gouvernance discipline le comportement du Léviathan (*théorème 1*) et génère une amélioration de Pareto dans la gestion des ressources naturelles (*théorème 2*). Ces théorèmes ont été étayés par les évidences tirées des données EPIN de la Banque mondiale. La base de données des pays exportateurs de carburant a été utilisée pour les analyses de robustesse.

**Mots-clés** : Ressources naturelles, Institutions, Leadership, Gouvernance, Groupe de pression.
**Classification JEL** : D78, E61, N5, Q3.

**Abstract** (*The artefact of the Natural Resources Curse*). This paper reexamines the validity of the natural resource curse hypothesis, using the database of mineral exporting countries. Our findings are as follows: (i) Resource-rich countries (RRCs) do not necessarily exhibit poor political, economic and social performance; (ii) RRCs that perform poorly have a low diversified exports portfolio; (iii) In contrast, RRCs with a low diversified exports portfolio do not necessarily perform poorly. Then, we develop a model of strategic interaction from a Bayesian game setup to study the role of leadership and governance in the management of natural resources. We show that an improvement in the leadership-governance binomial helps to discipline the behavior of lobby groups and generate a Pareto improvement in the management of natural resources. Evidence from the World Bank Group's CPIA data confirms the later finding. Our results remain valid after some robustness checks.

---

[*] Le premier draft de ce working paper a été élaboré par les auteurs à la suite d'une série de discussions durant leur séjour à la cité du fleuve, à l'université Mapon; Kindu, Maniema, RDC. Le contenu du présent papier n'engage que ses auteurs et aucunement leurs institutions d'attache respectives.
[†] Professeur à la Faculté d'Administration des affaires et Sciences économiques, Université Protestante au Congo, Kinshasa, RDC. Courriel: matataponyo@congochallenge.cd.
[‡] Ph.D. candidate et chargé de cours au Département des sciences économiques, Université du Québec, Montréal, CANADA. Bureau: DS-5820. Fax : +1 514-987-8494. Courriel: tsasa.jean-paul@uqam.ca.

# 1. Introduction

La littérature économique consacrée à l'analyse de l'hypothèse de la malédiction des ressources naturelles établit que les pays richement dotés en ressources naturelles, notamment en ressources minières, connaissent une croissance économique plus lente que ceux qui n'en disposent pas en abondance (*e.g.* le Japon, le Singapour ou la Corée du Sud)[1]. Plusieurs arguments, basés sur des évidences empiriques et théoriques, ont été développés pour expliquer l'hypothèse de la malédiction des ressources[2], notamment : (i) la volatilité du prix des matières premières (van der Ploeg et Poelhekke 2011, Moradbeigi et Law 2016); (ii) l'existence d'institutions oligarchiques, autocratiques ou dictatoriales (de Soysa 2002; Fearon 2005; Robinson et al. 2006, 2014; Kaldor et al. 2007; Botlhole et al. 2012; Wright et al. 2015; Caselli et Tesei 2016); (iii) l'existence d'institutions anarchiques avec non-application des droits de propriété (Beblawi 1987, Karl 1997, Auty 2001, Ross 2001, Chaudhry 2015); (iv) un épuisement rapide et non soutenable des ressources (Bhattacharyya et Collier 2014); (v) l'émergence de guerres civiles (Collier et Hoeffler 2000, Ross 2004, Collier et Hoeffler 2009, Koubi et al. 2009); (vi) une expansion cyclique du secteur des biens non échangeables via l'existence du syndrome hollandais[3] (Corden et Neary 1982, Krugman 1987, Ross 1999, Papyrakis et Gerlagh 2004, Slaibi et Kyle 2007); (vii) l'existence des rentes minières (Bates 2009, Marysse et Tshimanga 2013, Auty et Furlonge 2019) ou captation de rente (Boucekkine et Bouklia-Hassane 2011); (viii) le pouvoir de négociation asymétrique ou l'impossibilité de s'engager dans une planification à long terme (Humphreys et al. 2007).

Toutefois, de ce qui précède, deux remarques méritent d'être mises en évidence. D'une part, tous les pays riches en ressources naturelles n'affichent pas forcément des performances économiques, politiques ou sociales médiocres, à l'instar de la Norvège, des États-Unis ou du Canada. D'autre part, tous les pays sans ressources naturelles exportables ou abondantes (par exemple, le Haïti) n'atteignent pas évidemment les niveaux de croissance et de développement comparables à ceux des économies industrialisées telles que le Japon ou la Corée du sud. En même temps, lorsque l'on considère l'ensemble des pays riches en ressources naturelles, il advient que certains perform mieux à la fois économiquement, politiquement et socialement (Botswana, Chili, Malaisie), alors qu'en même temps d'autres affichent des performances médiocres. C'est notamment le cas, considérant la période 1960-2018, de la République démocratique du Congo (RDC), du Libéria et du Venezuela.

---

[1] Par exemple, Maddison (1991) note que le revenu par habitant de l'Australie est inférieur à celui du Japon, alors que ses ressources par habitant (calculées en fonction de la superficie occupée par le pays) sont 150 fois supérieures à celles du Japon. Sachs et Warner (1995, 1999) ont montré que les économies présentant un ratio élevé d'exportations de ressources naturelles par rapport au PIB en 1971 avaient tendance à avoir des taux de croissance faibles au cours de la période suivante (1971-1989).

[2] Auty (1993) définit la malédiction des ressources naturelles, ou *paradoxe de l'abondance*, comme les effets pervers de la richesse des ressources naturelles d'un pays sur son bien-être économique, social ou politique. Au mieux de notre connaissance, ce terme aurait été utilisé, pour la première fois par Alan H. Gelb, en référence aux propos, ci-après, de Pérez Alfonzo tenus en 1976 : « Dans dix ans, dans vingt ans, vous verrez que le pétrole nous apportera la ruine [...]. Ce sont les excréments du diable » (Gelb 1988, p. 321). En revanche, la première théorie de la malédiction des ressources naturelles remonterait à North et Thomas (1973). Ces derniers ont proposé un cadre d'analyse précurseur de la notion d'État rentier. Pour une revue de la littérature sur la malédiction des ressources, voir notamment Ross (1999, 2015), Rosser (2006), Humphreys et al. (2007), Di John (2011), van der Ploeg (2011), Frankel (2012), Morrison (2013), Havranek et al. (2016), Acar (2017), Badeeb (2017) et Arezki et al. (2018).

[3] Syndrome hollandais ou mal hollandais (en anglais *Dutch Disease*), terme utilisé pour la première fois par *The Economist* en 1977. URL : economist.com/the-economist-explains/2014/11/05/what-dutch-disease-is-and-why-its-bad.

Figure 1 : Illustration de l'artefact de la malédiction des ressources naturelles

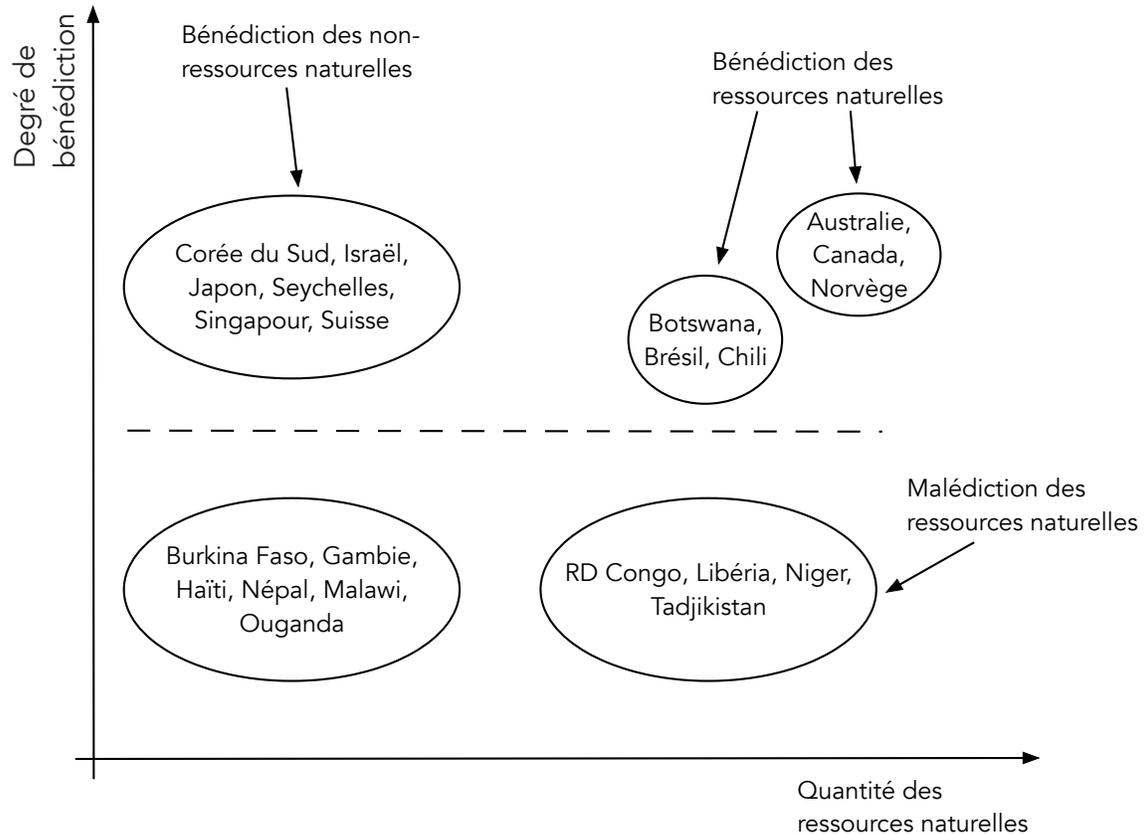

La figure 1 présente, pour un échantillon donné de pays, le produit cartésien de l'abondance relative des ressources naturelles par le niveau du développement économique relatif. Dans la zone supérieure de la figure: (i) les deux ovales à droite accueillent les pays abondamment dotés en ressources naturelles et ayant un niveau du développement élevé; (ii) l'ovale à gauche accueille les pays faiblement dotés en ressources naturelles et ayant un niveau du développement élevé. Dans la zone inférieure de la matrice: (iii) l'ovale à droite accueille les pays abondamment dotés en ressources naturelles et ayant un niveau du développement faible; (iv) l'ovale à gauche accueille les pays faiblement dotés en ressources naturelles et ayant un niveau du développement faible. Pour plus de détails sur la catégorisation des pays riches en ressources naturelles, cf. Auty (1993), BAD (2008), IMF (2012), Venables (2016), NationMaster (2019).

En parallèle, remarquons que les principales ressources pour le Seychelles sont la noix de coco, la cannelle, le poisson, les porcs, le sel et le fer. Pour la Gambie, ce sont le millet, les arachides, le poisson et les bovins. Ces deux pays ne recèlent donc pas de minéraux importants ou d'autres ressources naturelles. En sus, leurs bases agricoles respectives sont très limitées. Mais, entre 1960 et 2018, le revenu national brut par tête a été, en moyenne, près de vingt fois plus élevé au Seychelles (10 200 USD) qu'en Gambie (514 USD).

De même, les minerais et autres ressources naturelles au Japon et au Népal sont très limités. Cependant, entre 1960 et 2018, le RNB par tête a été, en moyenne, près de soixante fois plus élevé au Japon (36 611 USD) qu'au Népal (606 USD). Comme cela est illustré dans la figure 1, il y a lieu de trouver plusieurs autres combinaisons et exemples contrastants, notamment: [Singapour; Israël; Corée du sud; Suisse] contre [Burkina Faso; Haïti; Ouganda; Malawi]. En parallèle, il est possible de refaire le même exercice que précédemment, mais cette fois-ci, avec les pays abondamment dotés en ressources naturelles; en contrastant notamment, d'une part, le Libéria, la RDC, le Niger, et le Tadjikistan (pays à faible revenu) et, d'autre part, l'Australie, le Canada, le Chili et la Norvège (pays à revenu élevé) ou encore le Botswana et le Brésil (pays à revenu intermédiaire).

Ressort-il donc de la figure 1 que la malédiction des ressources n'est pas une évidence empirique absolue[4]. En réalité, la malédiction des ressources apparaît davantage comme une hypothèse, dont la validité théorique et empirique est à la fois relative et conditionnelle, plutôt qu'une règle inéluctable. D'où les questions que voici : pourquoi certains pays riches en ressources naturelles tirent profit de leurs dotations et d'autres non (*relativité*) ? Quels sont les facteurs qui sous-tendent la validité de l'hypothèse de la malédiction des ressources naturelles dans certains pays en développement (*conditionnalité*) ? À l'effet de répondre à ce questionnement, nous nous focalisons sur le cas des pays exportateurs de minerais et de métaux[5]. Par ailleurs, nous nous attachons à identifier les mécanismes potentiels susceptibles d'expliquer la persistance de la malédiction des ressources sur données dans certains pays en développement. En dernier lieu, nous discutons de la validité théorique et empirique de la malédiction des ressources naturelles dans les pays en développement et particulièrement en RDC. Nous soutenons et montrons à l'aide d'un modèle de jeux bayésiens que la malédiction des ressources n'est pas absolue. Elle est plutôt la résultante d'autres malédictions, principalement la « malédiction des institutions »[6], et plus spécifiquement la malédiction du leadership et de la gouvernance.

---

[4] Dans le même ordre d'idées, plusieurs études récentes, à la fois théoriques et empiriques, notamment celles de Brunschweiler (2008), Alexeev et Conrad (2009), Brunschweiler et Bulte (2009), Haber and Menaldo (2011), Kurtz et Brooks 2011, et Cotet et Tsui (2013) soulèvent des objections quant à la validité des recherches antérieures concernant la malédiction des ressources. Comme nous le verrons plus loin, la spécificité de notre papier est que nous considérons un mécanisme alternatif, basé sur le binôme leadership-gouvernance, pour montrer que les ressources naturelles ne sont pas fondamentalement une malédiction et, par corolaire, préciser dans quelle mesure elles peuvent donner l'impression d'être une malédiction.

[5] Dans ce papier, nous nous focalisons sur les ressources naturelles qui appartiennent à l'État, principalement les minerais et les métaux. Pour ces ressources, les rentes reviennent au secteur public et le gouvernement décide directement ou indirectement de la quantité de ressources à extraire. Par exemple, en RDC, le gouvernement décide de l'instruction cadastrale des demandes d'octroi, d'extension ou de renouvellement de mutation des droits miniers ou de carrières, cf. Décret no. 068/2003 du 03 avril 2003 portant statuts, organisation et fonctionnement du cadastre minier (CAMI). En référence au Larousse (2019), la rente dénote un revenu périodique, non obtenu par le travail, contrepartie du droit du propriétaire d'user de sa terre ou provenant d'une ressource rare.

[6] Le même argument peut être trouvé dans plusieurs autres documents, voir par exemple Mehlum et al. (2006), Robinson (2006), Tsui (2011) ou encore Sala-i-Martin et Subramanian (2013). Toutefois, contrairement aux études antérieures, dans ce papier nous insistons sur un type particulier d'institutions, à noter le binôme leadership-gouvernance (cf. Matata 2016). Le terme « malédiction des institutions »

Une caractéristique essentielle du modèle que nous développons est de tenir compte du fait que dans un pays riche en ressources naturelles comme la RDC, le décideur politique –en occurrence le premier ministre (ou son cabinet) ou le président de la République (ou son cabinet) – fait face à des groupes de pression puissants (McSparren et al. 2015, Fjeldstad et al. 2016) que nous qualifions de « Léviathan »[7]. En sus, nous supposons que le décideur est sujet à deux contraintes majeures, à noter : la contrainte de gouvernance et la contrainte de leadership (El Anshasy et Katsaiti 2013, Tsani 2013, Matata 2016). D'une part, la contrainte de gouvernance dénote l'incapacité du décideur politique à examiner chaque problème en profondeur et à analyser toutes les informations fournies par les groupes d'intérêts[8]. Ainsi, la contrainte de gouvernance peut obliger le décideur à prendre des décisions en se basant uniquement sur sa conviction *a priori* ou sur le signal du groupe de pression, comme nous l'illustrons dans les équilibres décrits dans le *lemme 1* (cf. Section 3)[9]. D'autre part, la contrainte de leadership traduit la limite supérieure du nombre des projets de réforme que le décideur est capable d'entreprendre avec succès[10]. De ce fait, la contrainte de leadership peut obliger le décideur à définir des priorités et donc choisir les questions à réformer et celles à laisser de côté (*lemme 2*). Ensuite, partant de ces considérations empiriques et analytiques, nous montrons qu'une amélioration du leadership peut aider à discipliner le comportement du Léviathan (*théorème 1*) et à générer une amélioration de Pareto (*théorème 2*) dans la gestion des ressources naturelles. En outre, sur base des données de la Banque mondiale, nous montrons que les pays riches en

---

est emprunté à Menaldo (2016, *chap. 4*). Voir Dietsche (2007), Weinthal et Luong (2006), et Stevens et Dietsche (2008) et Boucekkine et al. (2016) pour une interprétation alternative des interactions entre la dynamique des institutions et l'abondance relative des ressources naturelles. Voir aussi, en parallèle, Boucekkine et al. (2019) qui soutiennent que la version institutionnelle de la malédiction des ressources naturelles ne peut à elle seule expliquer l'hétérogénéité observée dans l'interaction entre la croissance et l'abondance relative en ressources naturelles.

[7] En référence aux écrits bibliques, le « Léviathan » dénote un monstre marin à plusieurs têtes (cf. Job 3:85, 40:25 et 41:16; Psaumes 74:142 et 104:263; Isaïe 27:14). Dans la culture générale, ce terme a été popularisé par Hobbes (1651), et récemment repris par Acemoglu et Robinson (2019) dans leur analyse sur la nature précaire de l'équilibre entre l'État et la Société. Dans notre papier, le terme Léviathan est défini comme une institution puissante (étatique, non étatique ou supra-étatique) qui contrôle un système donné. Cf. Section 2 pour plus de détails.

[8] Stevens et Dietsche (2008, p. 64) estiment que les recherches sur la malédiction des ressources naturelles devraient davantage consacrer d'efforts à l'exploration des conflits d'intérêts entre différents groupes sociaux. Voir aussi Guenther (2008) et Tsabora (2014).

[9] Dans la littérature économique, la gouvernance s'appréhende généralement comme les traditions et les institutions par lesquelles l'autorité d'un pays est exercée (Kaufman et al. 1999, Schneider 1999, Moore 2004, Iimi 2007, Edwards 2012, Kaufmann 2012, Asefa et Huang 2015, Rotberg 2015, Wolak 2016).

[10] Dans ce contexte, le leadership consiste à identifier et à éliminer les obstacles les plus contraignants à l'effet d'atteindre des objectifs spécifiques avec efficience (Zaleznik 1977, Maccoby 2000). Olson (1971) place le leadership au centre de l'économie politique, notamment dans son exposé sur les entrepreneurs politiques en tant qu'agents résolvant des problèmes d'action collective. Plus récemment, Jones et Olken (2005), Besley et al. (2016) et Matata (2016) ont également trouvé des preuves statistiquement significatives de l'importance du leadership dans l'explication des performances macroéconomiques. Voir Omotunde (2007), Collier (2008) et Myerson (2009, 2011a, b) pour les questions sur l'offre du leadership et la réputation du leader. Caselli et Cunningham (2007, 2009) décrivent une taxonomie des situations possibles qui déterminent les incitations des leaders. Voir aussi Auty and Gelb (2001) ou Murshed (2018).

ressources naturelles qui affichent des performances médiocres sont ceux qui exhibent un indice de leadership-gouvernance faible[11].

Le reste du papier s'organise comme suit. La section 2 présente quelques anomalies profondes dans la validité empirique de l'hypothèse de la malédiction des ressources naturelles. Nous y introduisons, par ailleurs, le concept de binôme leadership-gouvernance. La section 3 développe un modèle des jeux pour formaliser le rôle du binôme-gouvernance dans l'interaction entre le décideur politique et les groupes de pression dans la gestion des ressources naturelles. Ensuite, en utilisant la base des données EPIN du Groupe de la Banque mondiale, nous fournissons des preuves empiriques soutenant que la malédiction des ressources n'est qu'un artefact. Aussi, dans cette section, nous nous référerons à l'expérience singulière de la RDC pour montrer que le binôme leadership-gouvernance constitue un levier institutionnel particulier pour lever l'ambiguïté que pose souvent l'hypothèse de la malédiction des ressources naturelles dans les pays en développement. La section 4 procède à des vérifications supplémentaires afin de s'assurer de la robustesse des résultats empiriques établis dans la section précédente. La section 5 conclut.

**2. Des incohérences dans l'hypothèse de la malédiction des ressources naturelles**

La figure 2 résume les performances économiques des pays riches en ressources naturelles, dont les exportations en minerais et en métaux représentent plus de 40 pour cent des marchandises[12]. Ensuite, nous comparons les performances de ces pays à celles des pays à revenu élevé, dont les bénéfices tirés du pétrole représentent moins de 0.01 pour cent du PIB[13]. En effet, près de 84 pour cent des pays à faible revenu sont des économies africaines (Banque mondiale, WDI, 2019). Au même moment, lorsqu'on regarde les pays riches en ressources naturelles, dont les exportations en minerais et en métaux représentent plus de 50 pour cent des marchandises, 56 pour cent d'entre eux sont des économies à faible revenu et les 44 pour cent complémentaires sont des pays à revenu intermédiaire de la tranche inférieure. En sus, près de 43 pour cent des pays riches en ressources naturelles, dont les exportations en minerais et en métaux représentent plus de 50 pour cent des marchandises, sont des économies africaines (Nations unies, Division des Statistiques, 2019).

---

[11] De ce fait, l'indice de leadership-gouvernance s'interprète comme le pendant empirique des contraintes de gouvernance et de leadership dans le modèle théorique. Un indice de leadership-gouvernance faible est donc synonyme des contraintes de gouvernance et de leadership sévères. Et à l'inverse, Un indice de leadership-gouvernance élevé s'appréhende comme synonyme des contraintes de gouvernance et de leadership moins sévères.

[12] Les pays où les mines représentent au moins 40% des recettes d'exportation sont qualifiés de riches en ressources minières dans Auty (1993). Ce dernier précise que les pays riches en ressources minières peuvent être regroupés en deux catégories principales. D'une part, les exportateurs de minerais et, d'autre part, les producteurs d'hydrocarbures (cf. Auty 1993, p. 3).

[13] Nous excluons, de notre échantillon, les pays à revenu élevé dont les bénéfices tirés du pétrole représentent plus de 0.01 pour cent du PIB, en vue de souligner les performances économiques non expliquées fondamentalement par les ressources pétrolières, et se concentrer sur le différentiel induit par l'exploitation de minerais et de métaux. Plus précisément, les minerais et les métaux comprennent les produits des sections 27 de la CTCI (engrais bruts, minéraux); 28 (minerais métalliques, matériaux de récupération); et 68 (métaux non ferreux). Plus loin dans le texte (cf. Section 3), nous prendrons en compte les données sur les pays exportateurs de carburant à l'effet de s'assurer de la robustesse des faits stylisés que nous mettrons en évidence dans le présent papier.

Figure 2 : Comparaison des performances économiques des principaux pays exportateurs des minerais et des pays à revenu élevé (En moyenne. Période : 1960-2018)

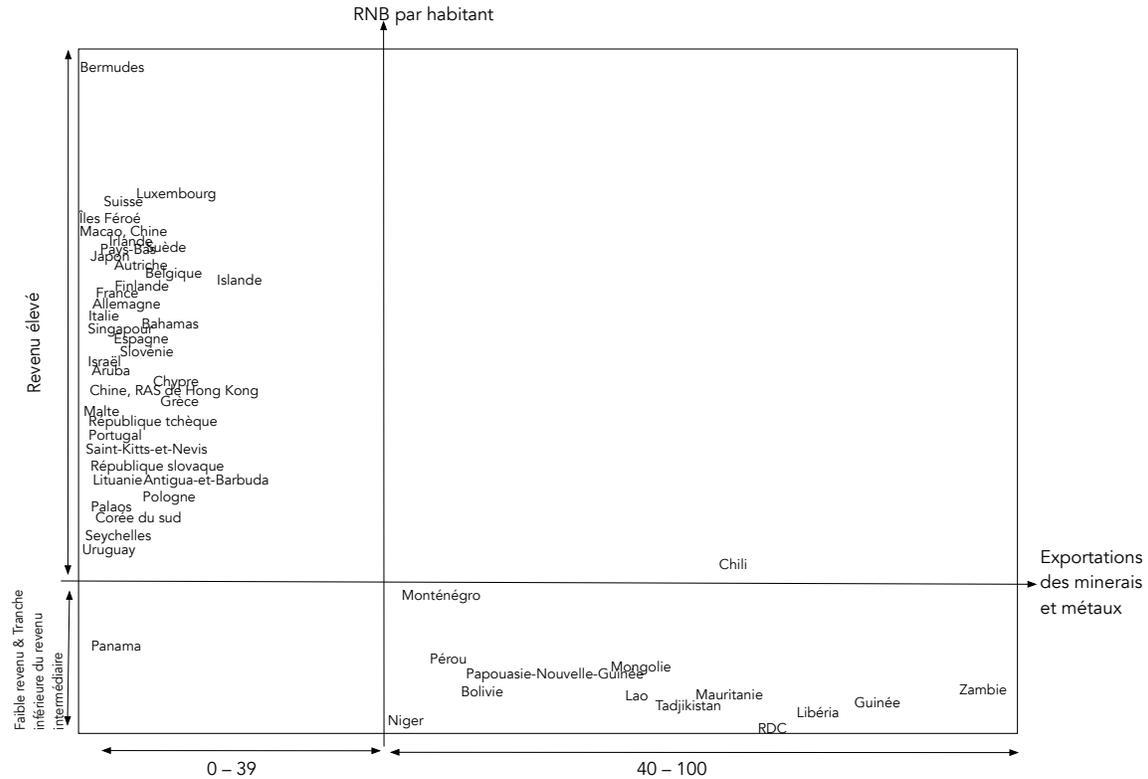

Note : La série sur le revenu national brut (RNB) par habitant provient des données des comptes nationaux de la Banque mondiale. La série sur les exportations de minerais et de métaux, mesurées en pourcentage des marchandises exportées, est tirée des estimations du personnel de la Banque mondiale issues de la base de données Comtrade maintenue par la Division des Statistiques des Nations Unies. La figure 1 se lit comme suit. Le cadrant supérieur droit regroupe tous les pays à revenu élevé dont les exportations en minerais et en métaux représentent plus de 40 pour cent des marchandises. Le cadrant supérieur gauche regroupe tous les pays à revenu élevé dont les exportations en minerais et en métaux représentent moins de 40 pour cent des marchandises. Le cadrant inférieur droit regroupe tous les pays à faible revenu ou à revenu intermédiaire de la tranche inférieure dont les exportations en minerais et en métaux représentent plus de 40 pour cent des marchandises. Le cadrant inférieur gauche regroupe tous les pays à faible revenu ou à revenu intermédiaire de la tranche inférieure dont les exportations en minerais et en métaux représentent moins de 40 pour cent des marchandises.

De la figure 2, il ressort clairement que, d'une part, les principaux pays exportateurs de minerais et de métaux appartiennent tous à la catégorie des économies à faible revenu ou à revenu intermédiaire de la tranche inférieure hormis le Chili et, d'autre part, les exportations de minerais et de métaux par les pays à revenu élevé, dont les bénéfices tirés du pétrole valent moins de 0.01 pour cent du PIB, représente en moyenne 3,4 pour cent des marchandises exportées. Autrement dit, à près de 90 pour cent, les pays riches en ressources naturelles, pour lesquels les minerais et les métaux composent plus de la moitié du portefeuille des exportations des marchandises, semblent vulnérables à la malédiction des ressources naturelles[14].

Figure 3 : Performances économiques des pays riches en ressources naturelles, producteurs de minerais et de métaux (En moyenne. Période : 1960-2018)

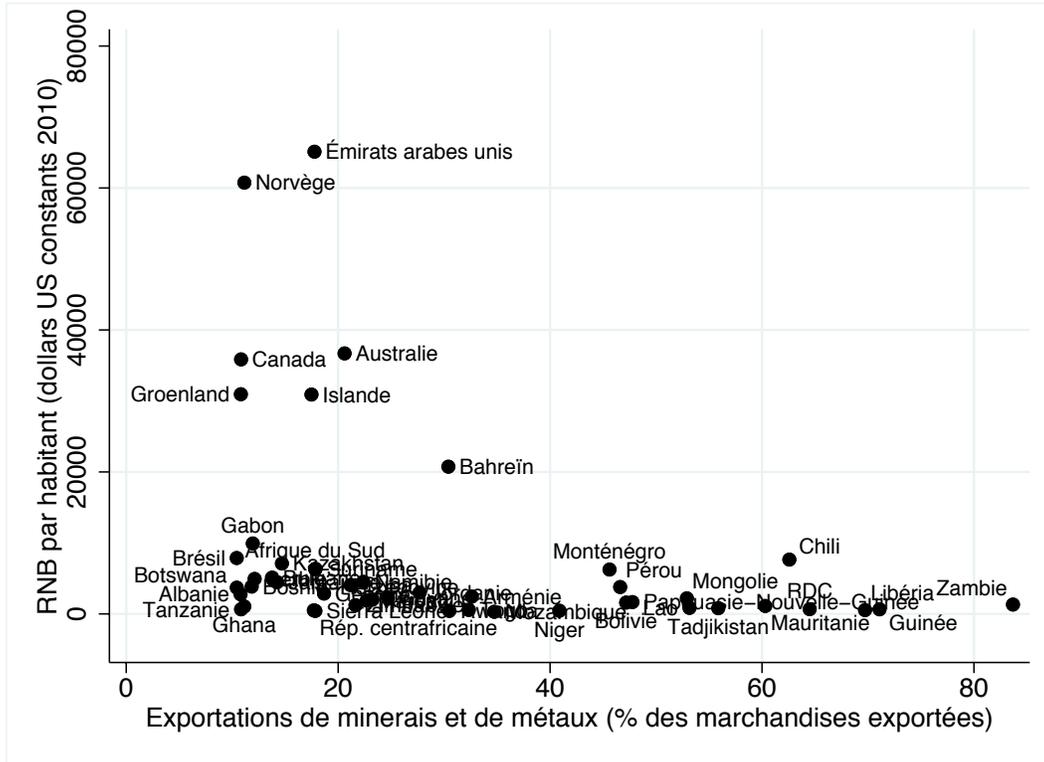

Note : Données des comptes nationaux de la Banque mondiale (RNB par habitant). Base de données Comtrade maintenue par la Division des Statistiques des Nations Unies (Exportations de minerais et de métaux). Cette figure établit la corrélation entre le niveau du développement économique et les exportations de minerais et de métaux. D'une part, plus un pays est localisé au nord du graphique, plus il a un niveau de revenu élevé. D'autre part, plus un pays est localisé à l'est du graphique, plus les minerais et les métaux représentent une part importante dans les exportations des marchandises. À l'inverse, d'une part, plus un pays est proche de l'axe des abscisses, plus il a un niveau de revenu faible. D'autre part, plus un pays est proche de l'axe des ordonnées, plus les minerais et les métaux représentent une part moins importante dans les exportations des marchandises.

Ces statistiques semblent appuyer la validité empirique de l'hypothèse de la malédiction des ressources naturelles. Cependant, en regardant de plus près la corrélation entre, d'une part, les performances économiques mesurées par le RNB par habitant et, d'autre part, les exportations de minerais et de métaux, il s'ensuit que la malédiction des ressources naturelles n'est pas spatialement absolue. En outre, en référence à la figure 3, l'hypothèse de la malédiction des ressources naturelles semble davantage être un « artefact » qu'une évidence empirique solide[15]. Autrement dit, la malédiction des ressources naturelles est en réalité une évidence empirique conditionnée par d'autres facteurs plus profonds.

---

[14] Ce constant fait de la « prépondérance des ressources naturelles dans le portefeuille des exportations » comme un symptôme du sous-développement induit par l'effet de l'abondance des minerais. Dans ce papier, nous réfutons cette corrélation fortuite et montrons qu'il existe des causes plus profondes permettant de mieux comprendre les raisons du sous-développement dans les pays riches en minerais.

En effet, la figure 3, en complément de la figure 1, montre que certains pays riches en ressources naturelles ont su véritablement tirer profit des recettes générées par les exportations de leurs minerais et métaux. C'est le cas notamment du Botswana, du Chili, du Monténégro et de la Norvège parmi tant d'autres. Cet ensemble inclut à la fois les pays en développement et les pays développés. Ici, nous conjecturons que le facteur profond qui permet de comprendre pourquoi certains pays riches en ressources naturelles tirent profit de leurs dotations et pourquoi certains autres n'en tirent pas, c'est le binôme leadership-gouvernance.

Le binôme leadership-gouvernance est un concept difficile à caractériser et à quantifier[16]. D'une part, le leadership suppose la prise des initiatives conformément à une vision, celle du leader, mais aussi cela suppose la mobilisation d'un sous-ensemble d'agents économiques œuvrant harmonieusement à l'effet d'atteindre des objectifs spécifiques (Zaleznik 1977, Maccoby 2000). D'autre part, la gouvernance peut être définie comme la manière dont les sociétés, les gouvernements et les organisations sont gérés et dirigés (Edwards 2012) ou comme les traditions et les institutions par lesquelles l'autorité d'un pays est exercée (Kaufman et al. 1999). Du point de vue de l'économie politique, la gouvernance fait référence à des éléments essentiels du vaste groupe d'institutions (voir Acemoglu et al. 2008, Fukuyama 2013). Les éléments particulièrement importants de la gouvernance incluent les institutions politiques d'une société (le processus de prise de décision collective et les contrôles exercés sur les hommes politiques et sur les groupes d'intérêts puissants sur le plan politique et économique), la capacité de l'État (la capacité à fournir des biens publics dans diverses régions du pays) et la réglementation des institutions économiques (comment l'État intervient pour encourager ou décourager l'activité économique de divers acteurs)[17].

Par définition, le leadership et la gouvernance sont donc des concepts multidimensionnels. Dans ce papier, nous nous intéresserons spécifiquement au binôme leadership-gouvernance dans le secteur public (i.e. dimension macroéconomique). Kaufmann et al. (1999) ont proposé un modèle permettant d'agréger des informations hétérogènes provenant de sources différentes pour produire des indicateurs globaux de gouvernance. Cependant, dans ce papier, notre attention se focalise non pas sur la gouvernance uniquement, mais sur le binôme leadership-gouvernance. À l'effet de caractériser et d'approximer la mesure de ce binôme, nous adoptons une définition qui se fonde sur ce que nous qualifiions de « Exigences de la gestion saine de la Res publica ».

---

[15] Plus loin, cf. Section 3, nous montrons plus rigoureusement que le fait, pour les pays en développement, de posséder les ressources naturelles n'est pas « absolument » une malédiction.

[16] En particulier, la difficulté vient en ce qu'un leader peut, par moment, tolérer le chaos et le manque de structure pour mieux comprendre les problèmes, éliminer les obstacles les plus contraignants et donc, atteindre la cible avec efficience (cf. Zaleznik 1977). Cette attitude, intrinsèquement assignée à tout leader, peut sembler contre-productive du point de la gouvernance en un instant donné du temps. Pour une discussion sur le leadership, voir House et Baets (1979), Peters (1967) et Hermalin (1998, 2013).

[17] Ainsi, Li (2005) note qu'une gouvernance de qualité implique une législation et un pouvoir judiciaire indépendants, des lois justes et transparentes avec une application impartiale, des informations financières publiques fiables et une grande confiance du public. Voir aussi Neo et Chen (2007).

Nous définissons le binôme leadership-gouvernance comme la capacité, pour un décideur politique, de garantir à la fois : un secteur financier efficient; la qualité de la gestion budgétaire; la mobilisation efficiente des revenus; l'efficacité des politiques structurelles; la qualité de l'administration publique; la stabilité du cadre macroéconomique; un niveau minimal acceptable de bien-être social; la transparence des activités dans le secteur public; un cadre politique favorisant le commerce de biens; les priorités nationales en matière de réduction de la pauvreté; la soutenabilité à court et à moyen terme de la politique fiscale; l'inclusion sociale et l'équité de l'utilisation des ressources publiques; la protection et l'utilisation durable des ressources naturelles et la gestion de la pollution; un environnement qui favorise la création d'emploi et une augmentation de la productivité; une structure de gouvernance basée sur le respect des droits de propriétés et sur des règles fiables; la promotion de l'égalité d'accès pour les hommes et les femmes à l'éducation, à la santé, à l'économie et à la protection; une stratégie de gestion de la dette favorisant la réduction des risques budgétaires et la soutenabilité de la dette à long terme; la fourniture de services par les secteur public et privé qui ont une incidence sur l'accès aux services d'éducation et de soins de santé et sur leur qualité.

En dépit de la pression provenant de divers groupes d'intérêt, il est de la fonction principale du leader exécutif de mobiliser tous les moyens nécessaires (en ressources humaines, matérielles, financières, psychiques et spirituelles) pour satisfaire en moyenne les exigences de la gestion saine de la Res publica susmentionnées. Nous nous référons à la base de données sur l'évaluation de la qualité des politiques et des institutions nationales (EPIN) du Groupe de la Banque mondiale (IDA 2017) pour quantifier chacune de ces dix-huit exigences.

Aussi convient-il de souligner que le binôme leadership-gouvernance a des conséquences sur la conception de la démocratie. En effet, puisque d'une part les politiciens ont leurs propres intérêts et que d'autre part, contrairement à la doctrine classique de la démocratie, la majorité ne représente pas forcément le peuple, Schumpeter (1942, p. 269) note avec fermeté que la démocratie ne doit plus être conçue comme une valeur suprême en soi, c'est-à-dire la volonté du peuple, mais surtout comme une concurrence pour le leadership politique[18]. De ce fait, le binôme leadership-gouvernance peut être véritablement effectif et donc socialement enrichissant, tout comme il peut être fondamentalement factuel et, en conséquence, socialement appauvrissant. En effet, le binôme leadership-gouvernance est véritablement effectif et socialement enrichissant si et seulement si le leadership et le mode de gouvernance sont compatibles à l'émergence d'un environnement institutionnel qui permet de bloquer ou contraindre le Léviathan[19]. Dans le cas contraire, le binôme leadership-gouvernance serait fondamentalement factuel et socialement appauvrissant.

---

[18] Cette assertion sur le rôle alternatif de la démocratie mérite une discussion plus approfondie à l'effet d'en apprécier l'opportunité et la pertinence. Toutefois, plus récemment, en utilisant une base de données de 1400 leaders mondiaux entre 1848 et 2004, Besley et Reynal-Querol (2011) trouvent que, comparativement aux autres régimes politiques, les démocraties ont environ 20% de chances supplémentaires de choisir des dirigeants très instruits. Pour une discussion sur le rôle de la démocratie sur la gouvernance des ressources naturelles en RDC, voir notamment Kabuya Kalala et Tshiunza Mbiye (2009) et Burnley (2011).

Le diable étant dans les détails (cf. Acemoglu et Robinson, 2019, chap. 9), il est important de noter que le binôme leadership-gouvernance ne consiste pas en la consécration d'un leader fort ou d'un monopole de la violence physique légitime incontesté (Weber 1917, 1919; Freund 1959). Il s'agit plutôt d'un mécanisme porté par un leader éclairé faisant usage d'une gouvernance de qualité devant garantir l'effet de la reine rouge (*Red Queen Effect*). Le binôme leadership-gouvernance contribue donc à l'émergence d'un cadre où le leader peut influer sur l'évolution des institutions[20], mais que son impact dépendra en grande partie de ces institutions et sera modulé par celles-ci en ce qu'il disposera de moins de pouvoir discrétionnaire pour mener des politiques qui seraient collectivement ou socialement désastreuses.

## 3. La malédiction des ressources naturelles : Relativité et conditionnalité

Dans ce papier, nous adressons deux questions principales. La première est celle de comprendre : *pourquoi certains pays riches en ressources naturelles tirent profit de leurs dotations et d'autres non ?* En effet soutenons-nous que si certains pays riches en ressources naturelles tirent profit de leurs dotations et d'autres non, c'est parce que dans la première catégorie des pays, les dirigeant font preuve de leadership « éclairé » et de gouvernance « de qualité ». Le leadership éclairé et la gouvernance de qualité se traduisent à leur tour par la promotion et le maintien des institutions pro-développement. La deuxième question que nous examinons est celle de préciser : *quels sont les facteurs qui sous-tendent la validité de l'hypothèse de la malédiction des ressources naturelles dans certains pays en développement ?* En effet, il y a plusieurs facteurs qui peuvent expliquer la validité de l'hypothèse de la malédiction des ressources naturelles dans certains pays en développement. Dans ce papier, nous défendons l'idée d'après laquelle la malédiction des ressources naturelles est uniquement valide dans les pays qui sont à la fois riches en ressources naturelles et dépourvus des dirigeants ayant la volonté et la capabilité nécessaires de négocier ou de résister aux lobbyistes ou groupes de pression[21]. Par exemple,

---

[19] Contraindre le Léviathan (*shackling the Leviathan*), terme emprunté à Acemoglu et Robinson (2019, p. 26). Le Léviathan est une institution étatique, non étatique ou supra-étatique puissante qui contrôle un système donné. Dans ce papier, le système fait essentiellement référence au secteur des ressources naturelles. Le blocage du Léviathan est le processus par lequel le public non élitiste de l'État obtient le contrôle des opérations de Léviathan en instituant des freins et contrepoids. Huntington (1993) note qu'une institution est efficace dès lors qu'elle peut limiter les excès des actions antagonistes.

[20] Comme ce fut le cas du rôle joué par Seretse Khama dans la mise en place des institutions pro-développement au Botswana ou du leadership de George Washington aux États-Unis.

[21] En effet, fondamentalement, le binôme leadership-gouvernance requiert non seulement la volonté mais aussi la capabilité. La volonté est comprise comme la disposition de caractère qui porte à prendre des décisions avec fermeté et à les conduire à leur terme en étant *potentiellement* prêt à surmonter tous les obstacles possibles. La capabilité dénote l'ensemble des ressources que devrait posséder le décideur politique afin de surmonter *effectivement* tous les obstacles possibles. Il s'agit notamment des ressources intellectuelles, morales, techniques, matérielles ou logistiques, financières, humaines et relationnelles. Prises séparément, « volonté » et « capabilité » ne sont que des conditions nécessaires, alors que réunies en couple, elles constituent une condition à la fois nécessaire et suffisante à la constitution stable du binôme leadership-gouvernance. Par exemple, le binôme leadership-gouvernance sous Lumumba en

McSparren et al. (2015) rappellent que dans la plupart des cas, les États africains n'ont pas la capacité technique d'exploiter les gisements de ressources sans s'associer à des sociétés multinationales qui fournissent à la fois les capitaux et l'expérience logistique. Par conséquent, en tant que producteur de produits non raffinés, les gouvernements africains sont chargés de renforcer la capacité de négociation de contrats avantageux avec les multinationales. Par ailleurs, plusieurs évidences empiriques et études dans la littérature, voir par exemple Fjeldstad et al. (2016), montrent que les lobbyistes ou groupes de pression peuvent mobiliser suffisamment des ressources (techniques, financières ou diplomatiques) pour promouvoir leur position auprès des hauts fonctionnaires dans les pays riches en ressources naturelles et donc influer sur les décisions politiques[22]. En conséquence, dans ce contexte, le problème n'est pas véritablement l'abondance des ressources naturelles[23], mais surtout le déficit des compétences ou du « contrepoids » nécessaire pour faire face aux lobbyistes ou groupes de pression dans la gestion des ressources naturelles.

Analytiquement, la meilleure façon de traiter cette problématique consiste à considérer un modèle de théorie des jeux où d'une part nous avons le décideur politique (DP), en occurrence le chef du gouvernement (ou son cabinet) ou le président de la République (ou son cabinet) et, d'autre part, les lobbyistes (*GI*). Plus spécifiquement, dans ce papier, nous considérons un jeu bayésien[24]. Le jeu se déroule dans un pays riche en ressources naturelles. La capacité de ce pays à bénéficier positivement de ses ressources naturelles dépend de la nature des choix du DP. Le DP fait face à diverses pressions de la part des groupes d'intérêt. Ses choix sont soumis à deux contraintes, à noter, la contrainte de gouvernance et la contrainte de leadership. Plus formellement et sans perte de généralité,

---

RDC était davantage caractérisé par la volonté que par la capabilité. Dans ce contexte, le binôme leadership-gouvernance ne pouvait être durablement stable dans le temps à cause du déficit de capabilité. En revanche, le binôme leadership-gouvernance sous la dictature du Général Park a permis l'émergence des institutions pro-développement car ce dernier avait su adjoindre, plus ou moins malicieusement, à sa volonté la capabilité. Cette dernière se traduit notamment par la normalisation des relations de la Corée du sud avec le Japon (cf. Traité nippo-sud-coréen du 22 juin 1965), la mise en place d'un service secret, la Korea Central Intelligence Agency, avec plus de quarante mille employés réguliers et un million de correspondants (Savada et Shaw 1992) ou encore la mobilisation de l'aide économique américaine. Hormis certaines considérations d'ordre dictatorial (Bouc 2018), les analystes s'accordent sur le fait que sans à la fois la volonté et la capabilité de Park Chung-hee, on n'aurait certainement pas parlé à ce jour de « The Miracle on the Han River » en Corée du sud (cf. Byeong-cheon et Yi 2006). De même, parmi tant d'autres exemples, George Washington, considéré comme le père de l'Amérique, a également su concilier volonté et capabilité et, à cet effet, largement contribué à l'émergence des institutions pro-développement aux États-Unis (Stephenson et Dunn 1940, Marling 1988, Estes 2001, Middlekauff 2015). La même lecture peut être faite pour Lee Kuan Yew (Singapour) ou pour Nelson Mandela (Afrique du Sud). Voir Sparks (1996), Barr (2000), Neo et Chen (2007), Carlin (2008), Read (2010) ou encore Johnston (2019).

[22] Par exemple, Fernandez et Rodrik (2007) montrent qu'une réforme peut être rejetée par un groupe de pression, alors même qu'elle pourrait largement améliorer le bien-être de tous.
[23] Ainsi parlions-nous de l'artefact de la malédiction des ressources naturelles.
[24] Voir Harsanyi (1967, 1968a, 1968b, 1982, 2004) et Kajii et Morris (1997) pour une introduction aux jeux bayésiens, et Austen-Smith et Wright (1992, 1994, 1996), Ainsworth (1993, 2002), Ainsworth et Sened (1993), Austen-Smith (1993 1995, 1996), Lowery (2007), Gabel et Scott (2011) pour une série de discussions sur le lobbying. Le choix de cette classe de modèles se justifie par le souci de formaliser le rôle du leadership. Cf. Acemoglu et Robinson (2013) ou encore Acemoglu et Jackson (2015) pour une discussion sur le rôle du leadership dans les modèles de théorie des jeux.

nous supposons que le DP devra choisir une politique sur deux options possibles, indexées par $i = 1, 2$. Nous désignons une politique par $p = (p_1, p_2)$, où $p_i \in \{0, 1\}$ est la politique sur l'option $i$. La politique $p_i = 1$ correspond à la décision de mettre en œuvre un projet de réforme sur l'option $i$. La politique $p_i = 0$ correspond au maintien du statu quo dans la gestion des ressources naturelles (renoncer à mettre en œuvre un projet de réforme). Nous notons par N$\in \{0, 1\}$ le nombre maximal de projets que le premier ministre peut réformer. En sus, nous supposons qu'il existe deux états du monde possibles pour chaque option. Nous désignons l'état de l'option $i$ par $\theta_i \in \{0, 1\}$. L'état $\theta_i = 1$ correspond aux circonstances dans lesquelles le DP profite de la réforme du dossier $i$. L'état $\theta_i = 0$ correspond aux circonstances dans lesquelles le DP bénéficie du maintien du statu quo sur ce dossier. Les États sont indépendants d'un dossier à l'autre. Le premier état qui a été réalisé pour un dossier $i$, $\theta_i$, est inconnu, mais sa distribution est connue de tous, i.e. du décideur politique et des groupes d'intérêt :

$$\theta_i = \begin{cases} 1, & \forall \pi_i \in \left(0, \frac{1}{2}\right) \\ 0, & \forall 1 - \pi_i. \end{cases} \quad (1)$$

L'utilité que le DP obtient de la politique $p = (p_1, p_2)$ dans l'état $\theta = (\theta_1, \theta_2)$ est donnée par :

$$U(p, \theta) = \alpha \times u_1(p_1, \theta_1) + u_2(p_2, \theta_2), \quad (2)$$

où $\alpha > 1$ représente l'importance du dossier 1 relativement au dossier 2.

Dans l'équation (2) :

$$u_i(p_i, \theta_i) = \begin{cases} 1, & \forall p_i = \theta_i \\ 0, & ailleurs \end{cases} \quad (3)$$

représente l'utilité du DP sous la politique $p_i$. Pour chaque dossier $i$, le DP préfère que la politique $p_i$ coïncide avec l'état réalisé $\theta_i$, ce qui donne au DP une information pertinente sur $\theta_i$. En outre, nous considérons qu'il y a deux groupes d'intérêt ($GI$), chacun plaidant pour une question distincte. Étant donné la politique $p = (p_1, p_2)$, le groupe d'intérêt défendant la question $i$ ($GI_i$) obtient l'utilité $v_i(p) = p_i$, ce qui signifie que le $GI_i$ cherche à maximiser la probabilité que $p_i = 1$, quel que soit l'état $\theta_i$. Considérant les évidences sur $\theta_i$, chaque $GI_i$ décide de faire pression ou non sur le DP. Si un $GI_i$ décide de faire du lobbying, il supporte un coût en services publics $f_i \in \{0, 1\}$. Si le DP lui en donne l'accès, il doit révéler $\theta_i$. Les coûts de lobbying, $f_i$, sont connus de tous. Le DP ne choisit qu'un seul $GI_i$. Comme nous l'avons évoqué précédemment, ce choix est réalisé sous contrainte de gouvernance et de leadership. La contrainte de gouvernance génère un coût d'opportunité endogène pour la société étant donné le choix effectif du DP. Le coût d'opportunité consiste à ce que le DP accorde l'accès à un $GI$ plutôt qu'à un autre. Ce coût d'opportunité peut donc être vu comme le gain social qu'implique le choix du DP. En parallèle, la contrainte de leadership, dans le cas d'espèce, tient à la capacité pour un DP à identifier et à réduire au minimum possible les obstacles les plus contraignants dans la gestion de la Res publica.

Étant donné la contrainte de gouvernance, nous considérons deux sous-jeux distincts. Un premier sous-jeu où le DP n'est pas contraint par la capacité à identifier et à réduire au minimum possible les obstacles les plus contraignants dans la gestion des questions-clés en lien avec la gestion des ressources naturelles. Dans ce cas, il peut examiner les deux options possibles, s'il le souhaite avant d'implémenter les réformes. Dans le deuxième sous-jeu, nous supposons que le DP est sujet à un déficit de leadership et ne peut considérer qu'une seule option dans le cadre de ses réformes. Le processus d'élaboration des politiques comporte quatre étapes.

Stade 1 : La Nature choisit l'état de chaque option $i$, et le révèle au groupe $GI_i$. La réalisation de l'état pour l'option $i$ est une information privée pour $GI_i$.

Stade 2 : Les $GI$ décident simultanément de faire pression ou non sur le Premier ministre (décideur politique, DP).

Stade 3 : Le Premier ministre observe le contenu de chaque dossier soumis par les $GI$, puis choisit à quel $GI$, le cas échéant, il accorde l'accès à la gestion des ressources naturelles. Une fois l'accès est accordé, le $GI_i$ doit révéler $\theta_i$ au DP.

Stade 4 : Le DP choisit la politique.

La solution dans ce jeu sera un équilibre bayésien parfait. Avant de procéder à la définition de l'équilibre, explicitons le processus d'élaboration des politiques.

*Stade 4 : choix de politique.* Au moment où le DP choisit sa politique, il a pris connaissance du contenu des dossiers soumis par les groupes de pression et de l'état réalisé pour l'option proposée par le $GI$ auquel l'accès avait été accordé, le cas échéant. Notons par $\lambda_i$ la décision du groupe de pression $GI_i$, telle que $\lambda_i = 1$ si $GI_i$ a exercé une quelconque pression et $\lambda_i = 0$ si non. Notons par $\psi_i$ la décision du DP à accorder l'accès au $GI_i$, où $\psi_i = 1$ si le DP a accordé l'accès à $GI_i$ et $\psi_i = 0$ si non. Compte tenu des profils de lobbying et de décisions d'accès, respectivement $\lambda = (\lambda_1, \lambda_2)$ et $\psi = (\psi_1, \psi_2)$, le DP forme la conviction $B_i(\lambda_i, \psi_i, \theta_i), \theta_i = 1$, en utilisant la règle de Bayes chaque fois que possible. Dès lors, une stratégie de politique formulée par le DP est telle que :

$$\rho: \{0,1\}^2 \times \{0,1\}^2 \to [0,1]^2 \qquad (4)$$

où $(\rho_1, \rho_2)(\lambda, \psi)$ dénote la probabilité que le DP choisisse la politique $p_1 = 1$ et la politique $p_2 = 1$, respectivement.

Pour $N = 2$, pour tout $i$, le DP maximise son utilité espérée avec la stratégie politique suivante :

$$\rho_i(\lambda, \psi) \begin{cases} = 1 & \text{si } B_i(\lambda_i, \psi_i, \theta_i) > \frac{1}{2} \\ \in [0,1] & \text{si } B_i(\lambda_i, \psi_i, \theta_i) = \frac{1}{2} \\ = 0 & \text{si } B_i(\lambda_i, \psi_i, \theta_i) < \frac{1}{2}. \end{cases} \quad (5)$$

Pour $N = 1$, pour tout $i$, le DP maximise son utilité espérée avec la stratégie politique suivante :

$$\rho_i(\lambda, \psi) \begin{cases} = 1 & \text{si } B_i > \frac{1}{2} \text{ et } \left(B_i - \frac{1}{2}\right) \times \alpha_i > \left(B_{-i} - \frac{1}{2}\right) \times \alpha_{-i} \\ \in [0,1] & \text{si } B_i \geq \frac{1}{2} \text{ et } \left(B_i - \frac{1}{2}\right) \times \alpha_i > \left(B_{-i} - \frac{1}{2}\right) \times \alpha_{-i} \\ = 0 & \text{ailleurs,} \end{cases} \quad (6)$$

avec $p_1 + p_2 \leq 1$, $\alpha_1 = \alpha$ et $\alpha_2 = 1$. De ce fait, lorsqu'il n'y a pas de contrainte de leadership ($N = 2$), le DP adopte la réforme sur la question $i$ s'il croit que $\theta_i = 1$ est plus probable que $\theta_i = 0$. En parallèle, lorsqu'il y a une contrainte de leadership ($N = 1$), le DP applique la réforme à l'option $i$, s'il croit que $\theta_i = 1$ est plus probable que $\theta_i = 0$ et, en outre, le gain d'utilité attendu de la réforme du problème $i$ dépasse celui de la réforme de l'autre. Ce dernier implique que lorsque $N = 2$ le choix politique du Premier ministre sur la question $i$ dépend de sa conviction, et donc de ses informations, sur $\theta_i$ uniquement, tandis que lorsque $N = 1$, cela dépend de ses convictions sur $\theta_1$ et $\theta_2$.

*Stade 3 : Accord de l'accès à la gestion des ressources.* Le DP ne peut accorder l'accès à un $GI$, que si ce dernier a choisi de faire pression sur lui. Au moment où le DP prend sa décision d'accès, il a déjà observé le contenu du dossier de chaque groupe de pression $GI_i$. Étant donné le dossier du groupe $GI_i$ en matière de lobbying, le premier ministre se fonde sur une conviction $B_i^A(\lambda_i)$ que $\theta_i = 1$, où l'exposant $A$ dénote le stade d'accès. Dès lors, une stratégie d'attribution d'accès par le DP est telle que :

$$\phi : \{0, 1\}^2 \rightarrow [0,1]^3 \quad (7)$$

où $\phi(\psi; \lambda)$ dénote la probabilité que le DP choisisse le profil d'accès $\psi = (\psi_1, \psi_2) \in \{(1,0), (0,1), (0,0)\}$ à partir du profil des décisions de lobbying $\lambda = (\lambda_1, \lambda_2)$.

Étant donnée la pression provenant des groupes d'intérêt $GI_i$, le DP choisit une stratégie d'accès qui résout le problème suivant :

$$\max_{(\phi_1, \phi_2) \in [0,1]^2} \mathbb{E}U(\phi_1, \phi_2 | \rho) \quad (8)$$
$$\text{sujet à : } \phi_1 + \phi_2 = 1,$$

tel que :

$$\phi_1 = \phi\big((1,0); (1,1)\big), \tag{9}$$

$$\phi_2 = \phi\big((0,1); (1,1)\big), \tag{10}$$

et où l'utilité espérée du PM au stade de l'accès est telle que :

$$\mathbb{E}U(\phi_1, \phi_2|\rho) \equiv \sum_{i=1}^{2} [\phi_i \times W_i(\rho) + (1-\phi_i) \times Z_i(\rho)] \times \alpha_i, \tag{11}$$

avec $W_i(\rho)$ la probabilité avec laquelle le DP pense choisir $p_i = \theta_i$ s'il donne l'accès à $GI_i$, et $Z_i(\rho)$ la probabilité avec laquelle le DP pense choisir $p_i = \theta_i$ s'il donne l'accès à $GI_{-i}$. Soit $X_i(\rho) \equiv W_i(\rho) - Z_i(\rho)$ la variation dans la probabilité que le DP choisisse $p_i = \theta_i$ en accordant l'accès à $GI_i$. Ensuite, la stratégie d'accès du DP lorsque chaque groupe d'intérêt fait pression est telle que :

$$\phi_i \begin{cases} = 1 & \text{si } X_i(\rho) \times \alpha_i > X_{-i}(\rho) \times \alpha_{-i} \\ \in [0,1] & \text{si } X_i(\rho) \times \alpha_i = X_{-i}(\rho) \times \alpha_{-i} \\ = 0 & \text{si } X_i(\rho) \times \alpha_i < X_{-i}(\rho) \times \alpha_{-i}, \end{cases} \tag{12}$$

où $\phi_1 + \phi_2 = 1$. Ainsi, chaque $GI$ font pression, le PM devra donner l'accès à celui qui dispose d'un dossier dont le bénéficie procure la plus grande valeur attendue pour la société.

*Stade 2 : Pression sur le décideur politique.* En tant que décideur politique, le premier ministre subit diverses pressions de la part des groupes d'intérêt. Ces groupes d'intérêt comprennent les lobbyistes internes et externes.

Une stratégie de lobbying pour un groupe d'intérêt donné $GI_i$ est telle que :

$$\xi_i : \{0, 1\} \to [0,1] \tag{13}$$

où $\xi_i(\theta_i)$ dénote la probabilité que $GI_i$ exerce une pression étant donné l'état $\theta_i$. Lorsque le groupe d'intérêt décide d'exercer une pression, il résout le problème suivant :

$$\max_{(\phi_1, \phi_2) \in [0,1]^2} \mathbb{E} p_i(\xi_i(\theta_i), \xi_{-i}, \phi, \rho) - \xi_i(\theta_i) \times g_i, \tag{14}$$

où $\mathbb{E} p_i(\xi_i(\theta_i), \xi_{-i}, \phi, \rho)$ est la probabilité que $p_i = 1$.

Nous disons que $GI_i$ fait une pression véridique si $\xi_i(\theta_i) = \theta_i$ pour chaque $\theta_i$, c'est-à-dire que $GI_i$ exerce une pression quand il détient une information favorable ($\theta_i = 1$) et s'abstient de faire du lobbying quand il a une information défavorable ($\theta_i = 0$). Nous disons que $GI_i$ fait un sur-lobbying si $\xi_i(1) = 1$ et $\xi_i(0) \in (0,1)$, i.e., $GI_i$ exerce une pression quand il dispose d'informations favorables et aléatoirement décide d'exercer ou non une pression quand il dispose d'informations défavorables.

*Stade 1 : Prospection.* La Nature choisit l'état de chaque option $i$, puis le révèle au groupe d'intérêt $GI_i$.

*Définition de l'équilibre.* À présent, nous pouvons définir l'équilibre. Dans ce jeu, un équilibre est un ensemble de stratégies $(\xi(\cdot), \phi(\cdot), \rho(\cdot))$ et un système de croyances $\left(B_i^A(\cdot), B(\cdot)\right)$ tels que l'ensemble de stratégies est séquentiellement rationnel compte tenu du système de croyances et les stratégies sont fondées sur la règle de Bayes chaque fois que cela est possible.

Examinons à présent les principales implications théoriques de notre modèle. Dans la section introductive, sur base des considérations empiriques, nous avons soutenu que la contrainte gouvernance peut obliger le décideur à prendre des décisions en se basant uniquement sur sa conviction *a priori* ou sur le signal du groupe de pression. Les équilibres décrits dans le lemme 1 formalisent cet argument.

**Lemme 1.** Soit N = 2-jeu.

1. Si :
$$\frac{\pi_1 f_1 + \pi_2 f_2}{\pi_1 \pi_2} \geq 1,$$
alors il existe un équilibre dans lequel $\lambda_i(\theta_i) = \theta_i$ pour chaque $\theta_i$ et chaque $i$.

2. Si :
$$\frac{\pi_1 f_1 + \pi_2 f_2}{\pi_1 \pi_2} < 1,$$
alors il existe un équilibre unique tel que pour $i = 1,2$ :

$$\begin{cases} \xi_i(1) = 1, \xi_i(0) = \frac{\pi_i}{1-\pi_i} \times \frac{1}{2\alpha_i - 1}, \text{ where } \alpha_1 = \alpha, \alpha_2 = 1 \\ \gamma_2 = 1 - \gamma_1 = \frac{f_1}{2\pi_2} \\ \rho_i(0,0) = 0, \rho_1(1,0) = 1, \rho_2(1,0) = \frac{\pi_2 \times (2\alpha - 1) \times f_2}{\pi_1 \times \alpha \times (2\pi_2 - f_1)}, \rho_i(1,1) = \theta_i \\ B_i(0,0; \theta_i) = B_i^A(0) = 0, B_1(1,0; \theta_i) = B_1^A(1) > \frac{1}{2} = B_2(1,0; \theta_i) = B_2^A(1). \end{cases}$$

Comme nous l'avons évoqué précédemment, $N$ est le nombre maximal de problèmes que le premier ministre peut réformer. Et donc, $N$-jeu dénote le jeu dans lequel le Premier ministre peut choisir de reformer un nombre illimité de questions, jusqu'à $N$. Dans le lemme 1, nous considérons le cas $N = 2$-jeu. Le choix de politique sur le dossier $i$ dépend ainsi de la croyance du DP sur ce dossier uniquement, $B_i(\lambda_i, \psi_i; \theta_i)$.

Par ailleurs, la contrainte de leadership peut obliger le décideur à définir des priorités et donc à choisir les questions à réformer et celles à laisser de côté.

**Lemme 2.** Soit N = 1-jeu.

1. Si $f_2 > 1 - \pi_1$, alors il existe un équilibre dans lequel les stratégies de lobbying sont données par :

$$\begin{cases} \xi_1(\theta_1) = \theta_1 \text{ pour chaque } \theta_1 \\ \xi_2(\theta_2) = 0 \text{ pour chaque } \theta_2. \end{cases}$$

   De plus, dans tout équilibre, les stratégies de lobbying sont données par :

$$\begin{cases} \xi_i(\theta_i) &= \theta_i \text{ pour chaque } \theta_i \\ \xi_{-i}(\theta_{-i}) &= 0 \text{ pour chaque } \theta_{-i}, \end{cases}$$

   pour certains $i \in \{1,2\}$.

2. Si $f_2 \leq 1 - \pi_1$, alors il existe un équilibre dans lequel $\xi_i(\theta_i) = \theta_i$ pour chaque $\theta_i$ et chaque $i$. Si, en outre, $f_i \leq 1 - \pi_i$ pour chaque $i$, les stratégies de lobbying à l'équilibre sont données par $\xi_i(\theta_i) = \theta_i$ pour chaque $\theta_i$ et chaque $i$.

Pour ne pas alourdir le texte, nous reprenons la preuve du lemme 1 et celle du lemme 2 en annexe. Une première implication importante de notre modèle est qu'une amélioration du leadership peut aider à discipliner le comportement des groupes d'intérêts. Le théorème 1 établit ce résultat.

**Théorème 1.**
1. Si $f_2 > 1 - \pi_1$, alors :

$$\begin{cases} \left|B_1^{N=2} - \frac{1}{2}\right| = \frac{1}{2} = \left|B_1^{N=1} - \frac{1}{2}\right| \\ \left|B_2^{N=2} - \frac{1}{2}\right| = \frac{1}{2} > \left|B_2^{N=1} - \frac{1}{2}\right|. \end{cases}$$

2. Si :

$$f_2 \in \left[\pi_1 \times \left(1 - \frac{f_1}{\pi_2}\right), 1 - \pi_1\right],$$

   alors :

$$\left|B_i^{N=2} - \frac{1}{2}\right| = \frac{1}{2} = \left|B_i^{N=1} - \frac{1}{2}\right|.$$

3. Si :

$$f_2 < \pi_1 \times \left(1 - \frac{f_1}{\pi_2}\right),$$

alors :
$$\left|B_i^{N=2} - \frac{1}{2}\right| \leq \frac{1}{2} = \left|B_i^{N=1} - \frac{1}{2}\right|,$$

avec une inégalité stricte pour certains $\theta$.

*Preuve.* La preuve du théorème 1 découle directement des lemmes 1 et 2. En effet, considérons premièrement le cas où $f_2 > 1 - \pi_1$. Une manipulation algébrique simple permet de réécrire cette inégalité comme suit :

$$\frac{\pi_1 f_1 + \pi_2 f_2}{\pi_1 \pi_2} > 1.$$

Pour $N = 2$-jeu, le lobbying à l'équilibre est véridique (lemme 1). Par conséquent, $B_i^{N=2} \in \{0,1\}$ pour chaque $i$. Pour $N = 1$-jeu, le groupe d'intérêt $GI_1$ fait pression, alors que $GI_2$ ne le fait pas (lemme 2). Par conséquent, $B_1^{N=1} \in \{0,1\}$ et $B_2^{N=1} = \pi_2$. À présent, considérons le cas où :

$$f_2 \in \left[\pi_1 \times \left(1 - \frac{f_1}{\pi_2}\right), 1 - \pi_1\right].$$

Autrement :

$$f_2 \geq 1 - \pi_1,$$

et

$$\frac{\pi_1 f_1 + \pi_2 f_2}{\pi_1 \pi_2} \geq 1.$$

Pour $N = 2$-jeu, le lobbying à l'équilibre et les croyances *a posteriori* sont telles que décrits dans le cas précédent. Pour $N = 1$-jeu, nous savons par le lemme 2 que le lobbying à l'équilibre est véridique. Par conséquent, $B_i^{N=1} \in \{0,1\}$ pour chaque $i$. Finalement, considérons le cas où :

$$f_2 < \pi_1 \times \left(1 - \frac{f_1}{\pi_2}\right).$$

Autrement :

$$f_2 < 1 - \pi_1,$$

et

$$\frac{\pi_1 f_1 + \pi_2 f_2}{\pi_1 \pi_2} < 1.$$

Pour $N = 1$-jeu, le lobbying à l'équilibre et les croyances *a posteriori* sont telles que décrits dans le cas précédent. Pour $N = 2$-jeu, nous savons par le lemme 1 que les groupes d'intérêt *GI*s font le sur-lobbying à l'équilibre. Par conséquent :

$$B_i^{N=2}(1,0; \theta_i) \in [1/2, 1)$$

pour chaque $i$. ∎

Le théorème 2 établit qu'une amélioration du leadership peut aider à générer une amélioration de Pareto.

**Théorème 2.** Soit $\mathbb{E}U_j^N$ l'équilibre *ex ante* espéré du gain du joueur $j \in \{1,2, DP\}$ dans le N-jeu. Il suit que $\mathbb{E}U_j^{N=1} \geq \mathbb{E}U_j^{N=2}$ pour chaque joueur $j \in \{1,2, DP\}$, avec au moins une inégalité stricte, si et seulement si :

$$\frac{\alpha \pi_1 f_1 + (2\alpha - 1) \times \pi_2 f_2}{\pi_1 \pi_2} \leq 1.$$

*Preuve*. Condition suffisante (« si »). Pour $\alpha > 1$ la condition

$$\frac{\alpha \pi_1 f_1 + (2\alpha - 1) \times \pi_2 f_2}{\pi_1 \pi_2} \leq 1$$

implique les inégalités strictes suivantes :

$$\frac{\alpha \pi_1 f_1 + \pi_2 f_2}{\pi_1 \pi_2} < 1$$

$$f_2 < 1 - \pi_1.$$

Ainsi, le deuxième cas du lemme 1 et le deuxième cas du 2 s'appliquent. Pour $N = 1$-jeu, le lobbying à l'équilibre est véridique et le DP donne priorité au dossier 1. En effet, le DP choisit $p = (1,0)$ lorsque $\theta = (1,1)$, sinon il choisit $p = \theta$. Dès lors, les gains espérés des joueurs sont tels que :

$$\begin{cases} \mathbb{E}U_1^{N=1} = \pi_1 \times (1 - f_1) \\ \mathbb{E}U_2^{N=1} = \pi_2 \times (1 - \pi_1 - f_2) \\ \mathbb{E}U_{DP}^{N=1} = \alpha + (1 - \pi_1 \pi_2). \end{cases}$$

Pour $N = 2$-jeu, les groupes d'intérêt $GI$s font le sur-lobbying à l'équilibre. Étant donné les stratégies et les croyances décrites dans la proposition (*théorème 2*) et dans la preuve du lemme 1, il suit que les gains espérés des joueurs se lisent comme suit :

$$\mathbb{E}U_1^{N=2} = \pi_1 \times (1 - f_1)$$

$$\mathbb{E}U_2^{N=2} = \pi_2 \times (1 - \pi_1 - f_2)$$
$$= \pi_2 \times \left\{ 1 - \pi_1 \times \left[ \frac{\alpha \times (2\pi_2 - f_1) - \frac{\pi_2}{\pi_1} \times (2\alpha - 1) \times f_2}{\pi_2 \times (2\alpha - 1)} \right] - f_2 \right\}$$

$$\mathbb{E}U_{DP}^{N=2} = (\alpha + 1) - \frac{2\pi_1 \pi_2 \alpha}{2\alpha - 1}.$$

En conséquence :

$$\mathbb{E}U_1^{N=1} = \mathbb{E}U_1^{N=2},$$

et

$$\mathbb{E}U_{DP}^{N=1} > \mathbb{E}U_{DP}^{N=2}.$$

En sus, la condition dans l'énoncé du théorème 2 implique :

$$\mathbb{E}U_2^{N=1} \geq \mathbb{E}U_2^{N=2}.$$

Condition nécessaire (« seulement si »). Supposons que :

$$\mathbb{E}U_j^{N=1} \geq \mathbb{E}U_j^{N=2}$$

pour chaque joueur $j \in \{1, 2, DP\}$, avec au moins une inégalité stricte. De ce fait, il faut avoir :

$$\frac{\pi_1 f_1 + \pi_2 f_2}{\pi_1 \pi_2} < 1. \tag{15}$$

Pour voir cela, supposons le contraire. Le premier cas du lemme 1 s'appliquerait alors. Ainsi, pour $N = 2$-jeu, le lobbying à l'équilibre serait véridique. Dans ce cas, le gain espéré du groupe de pression $GI_2$ à l'équilibre serait donc donné par :

$$\mathbb{E}U_2^{N=2} = \pi_2 \times (1 - f_2).$$

Pour $N = 1$-jeu, le groupe de pression $GI_2$ ferait un lobbying véridique chaque fois que :

$$f_2 \leq 1 - \pi_1,$$

ou s'abstiendrait de faire du lobbying lorsque :

$$f_2 > 1 - \pi_1.$$

Dès lors, le gain espéré du groupe de pression $GI_2$ à l'équilibre serait donné par :

$$\mathbb{E}U_2^{N=2} = \begin{cases} \pi_2 \times (1 - \pi_1 - f_2) & \text{si } f_2 \leq 1 - \pi_1 \\ 0 & \text{si } f_2 > 1 - \pi_1. \end{cases}$$

Ensuite, une manipulation algébrique simple permet d'établir que :

$$\mathbb{E}U_2^{N=2} > \mathbb{E}U_2^{N=1}.$$

D'où, la contradiction ! Nous pouvons finalement montrer que la condition (15) implique le deuxième cas du lemme 1, ainsi que le deuxième cas du lemme 2.

En effet, comme nous le savons de la preuve de la condition suffisance ci-dessus, il vient que :

$$\text{sign}\{\mathbb{E}U_2^{N=1} - \mathbb{E}U_2^{N=2}\} = \text{sign}\left\{\frac{\alpha\pi_1 \times f_1 + (2\alpha - 1)\pi_2 \times f_2}{\pi_1 \times \pi_2}\right\}.$$

D'où, la nécessité de cette condition car garantissant que l'inégalité (15) est satisfaite. ∎

Comme dans Robinson et al. (2006)[25], dans la suite de la présente section, nous fournissons quelques preuves empiriques pertinentes pour nos principaux résultats théoriques. Tout d'abord, nous montrons, en utilisant la base des données EPIN (évaluation de la qualité des politiques et des institutions nationales) du Groupe de la Banque mondiale (IDA 2017)[26], que les pays riches en ressources naturelles qui affichent des performances médiocres sont ceux qui exhibent un indice de leadership-gouvernance faible. Ensuite, en référence à l'expérience singulière de la RDC, nous montrons que le binôme leadership-gouvernance constitue un levier institutionnel particulier pour lever l'ambiguïté que pose souvent l'hypothèse de la malédiction des ressources naturelles dans les pays en développement et mettre en évidence le binôme leadership-gouvernance dans sa capacité à garantir la résilience économique face aux chocs sur les prix des matières premières.

En effet, dans la section 2, nous avons noté que la fonction principale d'un leader national, dans le sens du binôme leadership-gouvernance, est de mobiliser toutes les ressources nécessaires à l'effet de satisfaire en moyenne les exigences de la gestion saine de la Res publica. Pour satisfaire ces exigences, le décideur politique devrait souvent lutter contre les différents groupes de pression tant nationaux qu'étrangers et anticiper les tendances futures des fondamentaux de l'économie mondiale. Nous utilisons la moyenne des scores de l'ensemble de ces exigences telles que quantifiées dans la base des données EPIN pour approximer et quantifier l'indice du binôme leadership-gouvernance.

---

[25] Cf. Section 4 (Robinson et al. 2006, p. 462). Voir aussi Biresselioglu et al. (2019).

[26] En effet, le binôme leadership-gouvernance dénote la capacité, pour un décideur politique, de garantir : (1) la qualité de la gestion budgétaire; (2) un niveau minimal acceptable de bien-être social; (3) l'efficacité des politiques structurelles; (4) la promotion de l'égalité d'accès pour les hommes et les femmes à l'éducation, à la santé, à l'économie et à la protection; (5) la qualité de l'administration publique; (6) la stabilité du cadre macroéconomique; (7) la transparence des activités dans le secteur public; (8) un cadre politique favorisant le commerce de biens; (9) la soutenabilité à court et à moyen terme de la politique fiscale; (10) une structure de gouvernance basée sur le respect des droits de propriétés et sur des règles fiables; (11) la fourniture de services par les secteur public et privé qui ont une incidence sur l'accès aux services d'éducation et de soins de santé et sur leur qualité; (12) une stratégie de gestion de la dette favorisant la réduction des risques budgétaires et la soutenabilité de la dette à long terme; (13) un secteur financier efficient; (14) la protection et l'utilisation durable des ressources naturelles et la gestion de la pollution; (15) l'efficacité de la mobilisation des revenus; (16) un environnement qui favorise la création d'emploi et une augmentation de la productivité; (17) les priorités nationales en matière de réduction de la pauvreté; (18) l'inclusion sociale et l'équité de l'utilisation des ressources publiques. Bien évidemment, il est crucial de souligner que, par construction, le binôme leadership-gouvernance va donc bien au-delà des considérations se limitant uniquement à la gestion efficiente des ressources naturelles. La description des indices EPIN est reprise en Annexe A.3.

Figure 4 : Produit cartésien du binôme leadership-gouvernance par l'abondance relative en ressources naturelles

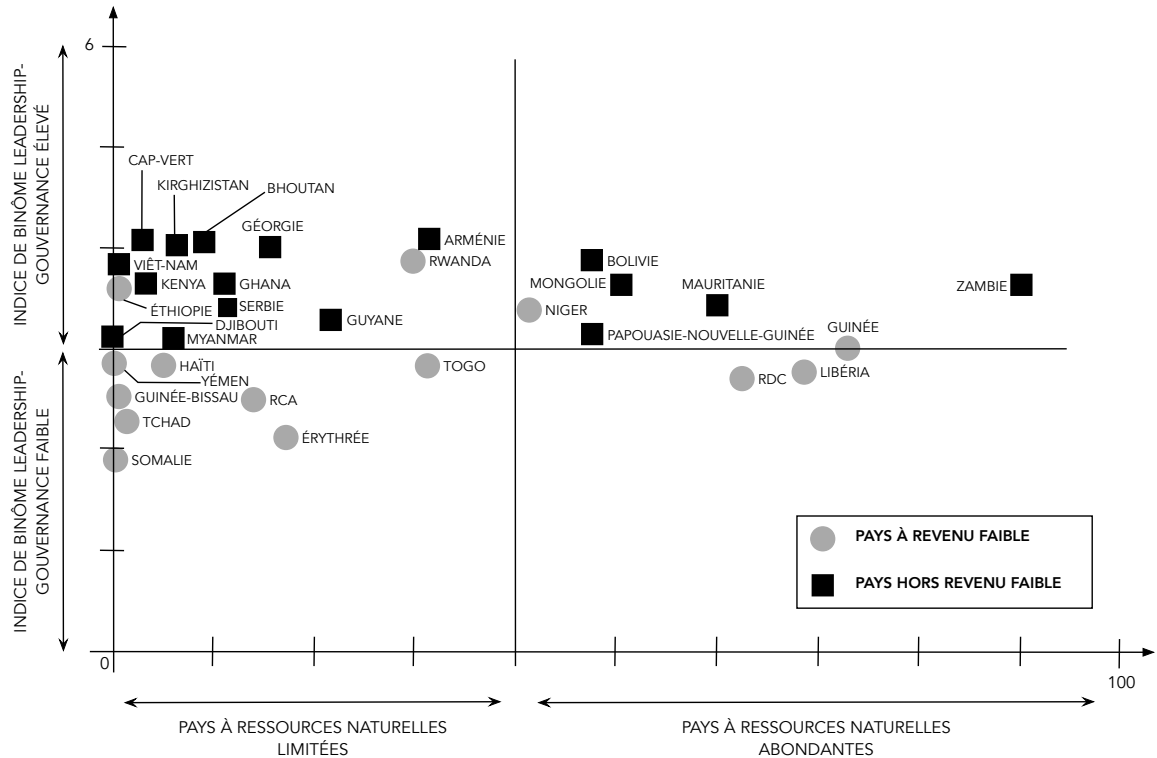

Note : Ce graphique mesure la corrélation entre le binôme leadership-gouvernance et l'abondance relative en ressources naturelles. Le binôme leadership-gouvernance mesure la moyenne des scores de l'ensemble de ces exigences (Période: 2005-2018) telles que quantifiées dans la base des données EPIN de la Banque mondiale (IDA 2017). Plus précisément, l'indice du binôme leadership-gouvernance est approximation quantitative du leadership éclairé et de la gouvernance de qualité (Classement de l'indice du binôme leadership-gouvernance : 1=faible et 6=élevé). L'abondance relative en ressources naturelles est mesurée par les exportations de minerais et de métaux en pourcentage des marchandises exportées (Période: 1960-2018). Suivant la classification de la Banque mondiale, un RNB par habitant inférieur ou égal à 1025 dollars américains définit les pays à faible revenu; cf. Banque mondiale (2020) pour plus de détails. La taille de l'échantillon a été conditionnée par la limitation et la disponibilité des données.

Trois observations principales émergent de la figure 4. Tout d'abord, il ressort que tous les pays hors revenu faible sont ceux qui exhibent un indice de binôme leadership-gouvernance relativement plus élevés. Quelques exceptions sont à noter à cet effet. C'est notamment le cas de l'Éthiopie, du Rwanda et du Niger. Ces trois pays exhibent un indice de binôme leadership-gouvernance élevé mais sont encore catégorisés dans le groupe des pays à revenu faible, i.e. économies dont le RNB par habitant, calculé à l'aide de la méthode dite de « l'Atlas », est de 1025 dollars américains ou moins en 2018. La deuxième observation tient à ce que l'abondance ou la rareté relative en ressources naturelles n'aide pas à comprendre pourquoi certains pays performent mieux et que d'autres non. En effet, au regard des évidences empiriques qui se dégagent de la figure 4, il advient que soutenir l'hypothèse de malédiction des ressources naturelles apparaît davantage comme un artefact car en réalité on peut toujours trouver des pays riches en ressources naturelles dont certains sont des économies à faible revenu et que d'autres sont des économies hors faible revenu[27]. La principale conclusion qui se conforme à la régularité des faits est que les pays hors revenu faible exhibent tous, sans exception, un indice de binôme leadership-gouvernance relativement plus élevé. La troisième observation, et la plus importante, est quelque peu le corollaire des deux premières observations. Elle concerne, en effet, la question de comprendre : *pourquoi certains pays riches en ressources naturelles tirent profit de leurs dotations et d'autres non ?* Très clairement, la figure 4 montre que si certains pays riches en ressources naturelles tirent profit de leurs dotations et que d'autres non, c'est parce que dans la première catégorie des pays, les dirigeant font preuve de leadership éclairé et de gouvernance de qualité. Autrement dit, il n'y a pas de malédiction des ressources naturelles, il n'y a que la défaillance du binôme leadership-gouvernance. Ainsi, contrairement à la figure 1 ou à la figure 2, la figure 4 permet d'apporter un support empirique plausible dans la compréhension de la question : « pourquoi certains pays riches en ressources naturelles tirent profit de leurs dotations et d'autres non ? »

La spécificité du mécanisme explicatif que nous proposons dans le présent papier, pour lever l'ambiguïté que pose la validité empirique de la malédiction des ressources naturelles, s'inscrit dans la lignée de celui proposé par Robinson et al. (2006), Kabuya Kalala et Tshiunza Mbiye (2009), Menaldo (2016), parmi tant d'autres. Toutefois, contrairement à la littérature précédente, nous insistons sur un type particulier d'institutions, à noter le binôme leadership-gouvernance. Comme cela est en partie documenté et soutenu par Stevens et Dietsche (2008, p. 64), et Parker et Vadheim (2017, p. 2), nous avons exploité dans notre modèle le fait que, dans la gestion des ressources naturelles, le décideur politique est confronté à des groupes de pression ou des groupes d'intérêt puissants (« Léviathan »). Et pour résister aux pressions diverses provenant de ces différents groupes d'intérêt, le décideur devra davantage faire preuve d'un leadership éclairé et d'une gouvernance de qualité.

L'indice que nous considérons pour évaluer quantitativement le binôme leadership-gouvernance est une moyenne arithmétique des scores des indices EPIN du groupe de la Banque mondiale. En regardant de plus près la composition des indices utilisés dans le calcul de cette moyenne, nous trouvons notamment l'indice de la protection et l'utilisation durable des ressources naturelles et la gestion de la pollution. D'emblée, la question naturelle qui émerge est celle de savoir si ce dernier indice peut aider à comprendre « pourquoi certains pays riches en ressources naturelles tirent profit de leurs dotations et d'autres non ? » Nous répondons à cette question dans la figure 5. En comparant la figure 4 et la figure 5, il ressort que l'indice du binôme de leadership-gouvernance (*indice B-LG*) semble plus discipliné que l'indice de l'utilisation durable des ressources naturelles (*indice U-RN*) dans la prédiction des performances économiques des nations; cf. Tableau 1.

---

[27] En effet, si l'hypothèse de la malédiction des ressources naturelle étaient empiriquement valide et plausible, tous les pays avec des ressources naturelles abondantes devraient être localisés dans le cadran IV, c'est-à-dire le cadran à droite dans le panel inférieur de la figure 4 (Sud-Est).

Figure 5 : Produit cartésien de l'utilisation durable des ressources naturelles par l'abondance relative en ressources naturelles

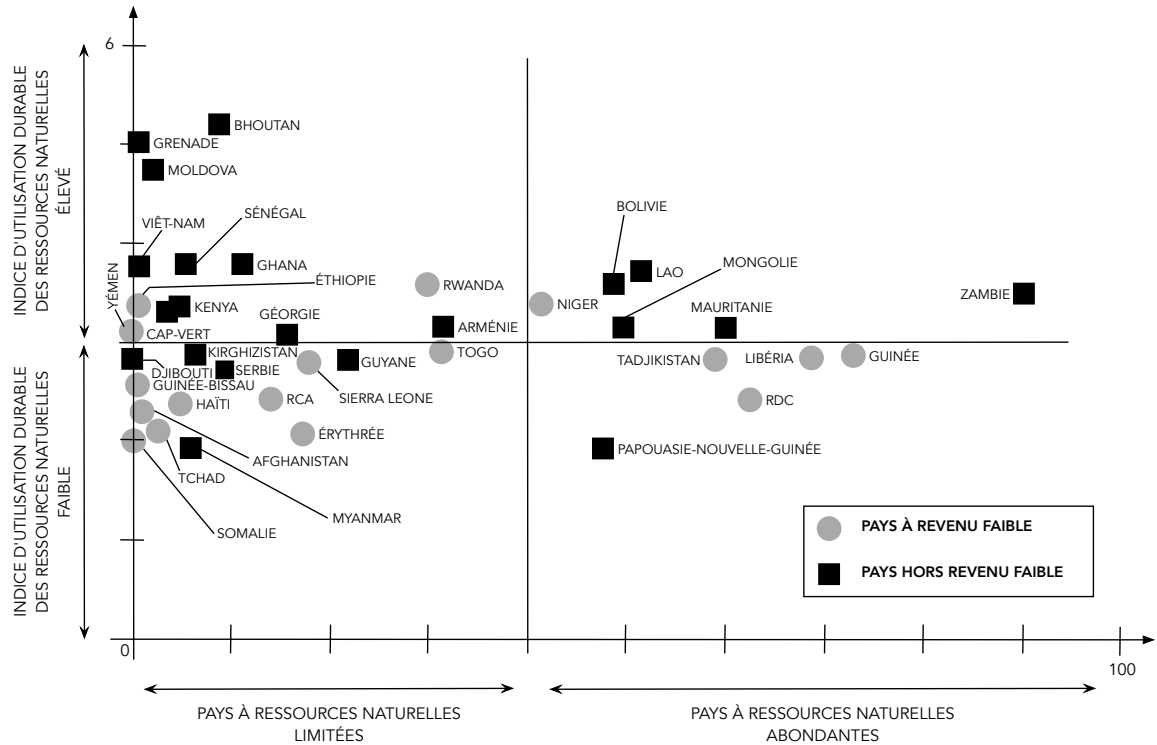

Note : Ce graphique mesure la corrélation entre l'utilisation durable des ressources naturelles et l'abondance relative en ressources naturelles. L'utilisation durable des ressources naturelles mesure la protection et l'utilisation durable des ressources naturelles et la gestion de la pollution (Classement : 1=faible et 6=élevé). L'abondance relative en ressources naturelles est mesurée par les exportations de minerais et de métaux en pourcentage des marchandises exportées (Période: 1960-2018). Suivant la classification de la Banque mondiale, un RNB par habitant inférieur ou égal à 1025 dollars américains définit les pays à faible revenu; cf. Banque mondiale (2020) pour plus de détails. La taille de l'échantillon a été conditionnée par la limitation et la disponibilité des données.

Tableau 1 : Comptabilité des anomalies dans la prédiction des performances économiques par les indices B-LG et U-RN

|  | Cadran I | Cadran II | Cadran III | Cadran IV | Total Nombre d'anomalies |
|---|---|---|---|---|---|
| Indice B-LG | 1 | 2 | 0 | 0 | 3 |
| Indice U-RN | 1 | 3 | 5 | 1 | 10 |

Note : Ce tableau propose une comptabilité des anomalies qu'exhibent, d'une part, le produit cartésien du binôme leadership-gouvernance par l'abondance relative en ressources naturelles (figure 4) et, d'autre part, le produit cartésien de l'utilisation durable des ressources naturelles par l'abondance relative en ressources naturelles (figure 5).

Figure 6 : Évolution du prix des matières premières sur les marchés mondiaux et des recettes publiques en République démocratique au Congo (RDC)

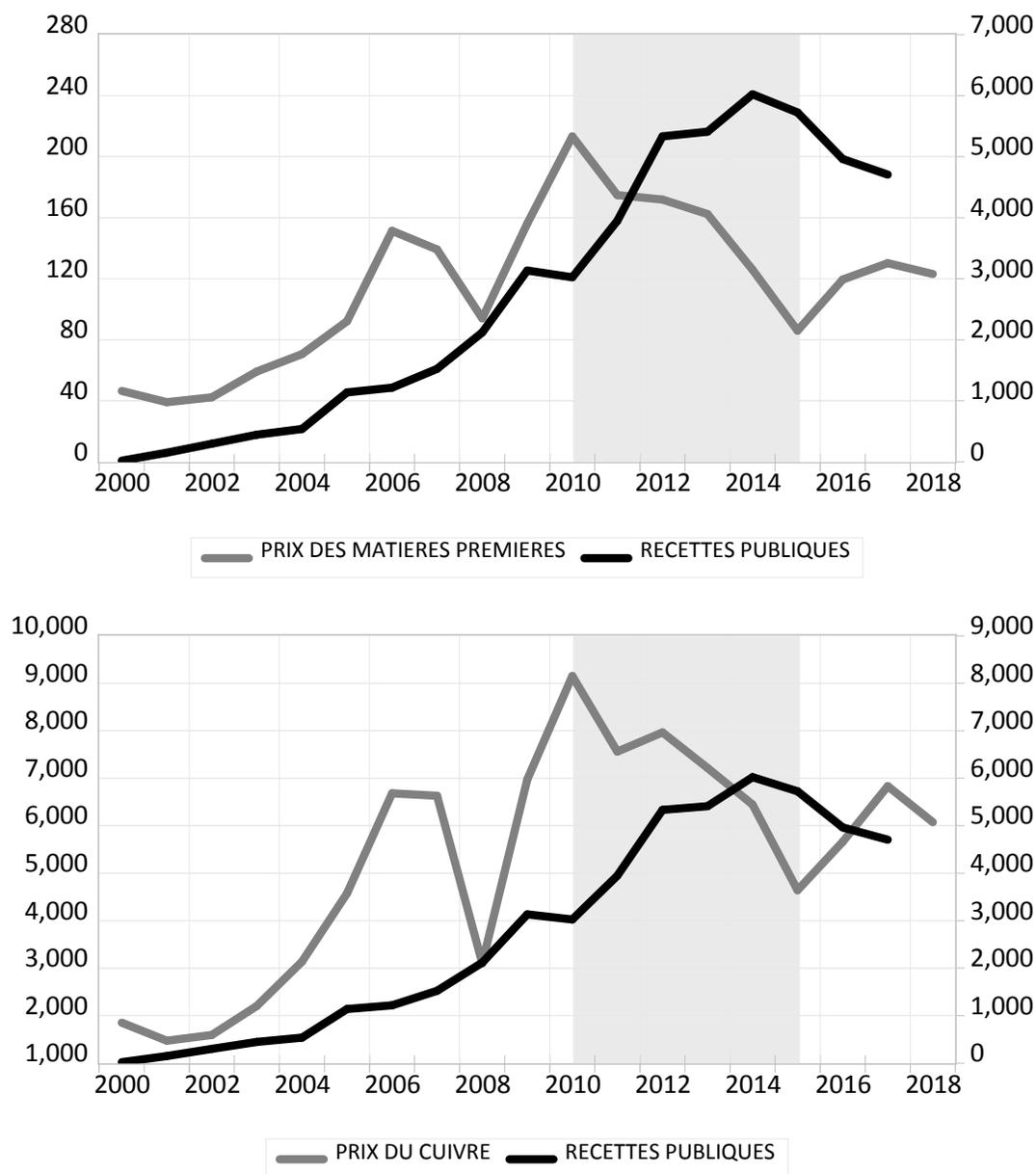

Note: Graphique réalisé sur base des statistiques macroéconomiques mondiales du FMI (IMF Cross-Country Macroeconomic Statistics) publiées par Open Data for Africa (2019). Le prix des matières premières correspond à l'indice des prix des métaux de base (Année de base 2016). Les métaux de base comprennent l'aluminium, le cobalt, le cuivre, le fer, le nickel, l'étain. Unité: Indices des prix de l'uranium et du zinc. Url: quandl.com/data/ODA/PMETA_INDEX. En parallèle, le prix du cuivre est évalué en dollar américain par tonne métrique. Url: quandl.com/data/ODA/PCOPP_USD. Les recettes publiques sont mesurées en milliards de Francs congolais. Les données sur les recettes publiques en RDC proviennent des rapports annuels de la Banque centrale du Congo (BCC 2010, p. 75; BCC 2017, p. 71). Pour une meilleure visualisation des chroniques, les recettes publiques sont évaluées dans l'*axe droit*, alors que le prix des matières premières et le prix du cuivre sont évalués dans l'*axe gauche* respectivement.

En effet, théoriquement, on s'attend à ce que : (i) tous les pays avec des ressources naturelles relativement abondantes exhibant un « indice » élevé soient localisés dans le cadran I (Nord-Est); (ii) tous les pays avec des ressources naturelles relativement limitées ayant un « indice » élevé soient localisés dans le cadran II (Nord-Ouest); (iii) tous les pays avec des ressources naturelles relativement limitées affichant un « indice » faible soient localisés dans le cadran III (Sud-Ouest); (iv) enfin, tous les pays avec des ressources naturelles relativement limitées exhibant un « indice » faible soient localisés dans le cadran IV (Sud-Ouest). Empiriquement, comme nous le résumons dans le tableau 1, il ressort que l'indice du binôme leadership-gouvernance performe mieux que l'indice de l'utilisation durable des ressources naturelles. Cette conclusion conforte davantage la discussion entreprise dans la note de page 21, celle de l'exigence d'avoir à la fois et non séparément un leadership éclairé et une gouvernance de qualité.

Pour clore la présente section, nous montrons à l'aide d'un exemple caractéristique qu'une économie peut, en dépit de sa dépendance en ses ressources naturelles, afficher une forte résilience face aux chocs des prix des matières sur les marchés mondiaux à condition que le décideur politique face preuve, à la fois, d'un leadership éclairé et d'une gouvernance de qualité. Nous formulons cette prédiction, autrement, dans la conjecture 1.

> **Conjecture 1.** *Lorsqu'un pays avec des ressources naturelles abondantes affiche une forte vulnérabilité face aux chocs des prix des matières sur les marchés mondiaux, cela serait symptomatique d'une défaillance du binôme leadership-gouvernance.*

Une illustration de cette conjecture est proposée dans la figure 6, où il se constate que durant la période 2011-2015, les recettes publiques en RDC étaient en hausse monotone malgré le fait que les prix des matières premières sur les marchés mondiaux, et plus particulièrement celui du cuivre, aient baissé constamment. Dans la discussion qui suit, nous montrons que cette performance s'inscrit dans un cadre plus global et qu'elle n'est pas le fait d'une bonne fortune mais plutôt la conséquence d'un leadership éclairé et d'une gouvernance de qualité. Pour ce faire, nous considérons les quatre derniers exécutifs, respectivement avec comme Chef du Gouvernement : Antoine Gizenga (2006-2008); Adolphe Muzito (2008-2012); Matata Ponyo (2012-2016); Badibanga/Tshibala (2016-2018) [28]. La figure 7 présente un ensemble de statistiques en considérant le mandat de chaque exécutif. En effet, comme nous pouvons le voir, la période 2012-2016 semble être singulière sur plusieurs dimensions, comparativement aux autres périodes : une croissance économique forte; un niveau général des prix contrôlé; un taux de change nominal stable; une politique monétaire et une gestion budgétaire accommodantes avec un taux d'intérêt directeur stable et un élargissement de l'assiette fiscale. *Est-ce que ces performances économiques sont le fait de la clémence des prix des matières premières sur les marchés mondiaux ou d'autres facteurs exogènes ?*

---

[28] La non disponibilité des données en haute fréquence nous contraint de considérer le gouvernement Badibanga et le gouvernement Tshibala comme un seul exécutif. Samy Badibanga fut Chef du gouvernement du 17 novembre 2016 au 18 mai 2017, alors que Bruno Tshibala l'a été du 18 mai 2017 au 7 septembre 2019. Avant leur nomination en qualité de premier ministre, Samy Badibanga et Bruno Tshibala étaient membres effectifs de l'Union pour la démocratie et le progrès social (UDPS).

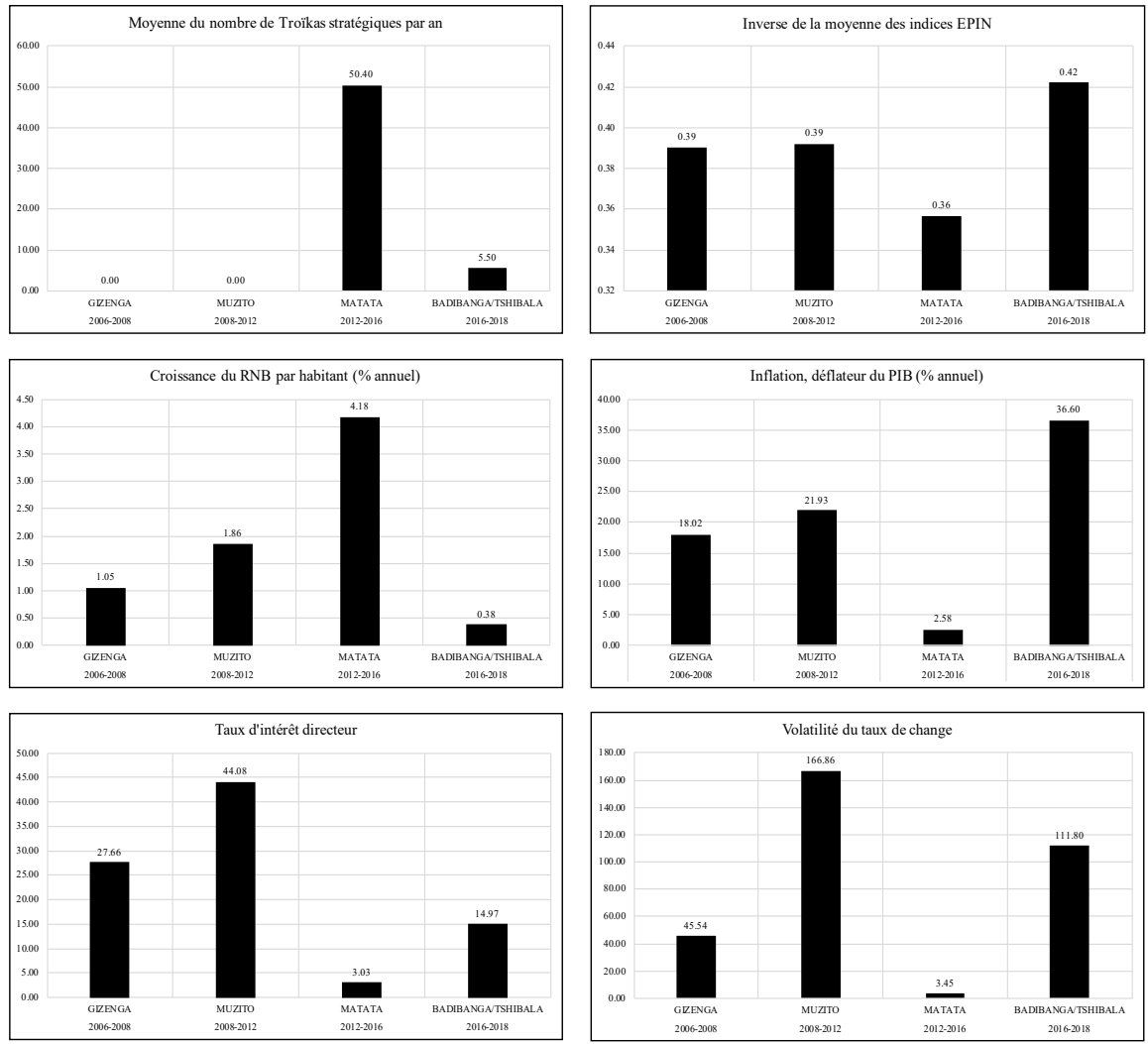

Figure 7 : Performances économiques, leadership et gouvernance en RDC

Note : La série sur le taux de croissance du revenu national brut (RNB) par habitant, du taux d'inflation et des indices EPIN (CPIA index) proviennent des données des comptes nationaux de la Banque mondiale (WDI). Les données sur la moyenne du nombre de Troïkas stratégiques proviennent du Cabinet du Premier ministre de la RDC.

En effet, entre 2012 et 2016, force est de constater qu'au-delà des mécanismes classiques du travail interministériel que sont les commissions interministérielles permanentes, le premier ministre avait initié une série de réformes institutionnelles en vue de créer les conditions favorables pour une amélioration durable du fonctionnement des institutions publiques. De ce fait, plusieurs plateformes d'interactions ont été mises en place, principalement la Troïka stratégique[29]. Remarquons tout de suite que l'instauration de la

---

[29] Le premier ministre Matata Ponyo, alors ministre des finances (2010-2012), avait également instauré en son temps un mécanisme quasiment identique dénommé Troïka politique. Contrairement à la Troïka stratégique (cf. Annexe A.3 pour détails), la Troïka politique se focalisait surtout sur la coordination des politiques monétaire et budgétaire, et des réformes pilotées par le ministère des finances.

Troïka stratégique et une nouvelle approche du management de la Res publica coïncident avec la période où les symptômes de la malédiction des ressources naturelles (pays riches en ressources naturelles avec un profil de croissance économique faible, une inflation et un taux de change volatiles) semblent véritablement céder la place aux symptômes de l'émergence économique[30]. Ce résultat conforte davantage l'idée d'après laquelle le binôme leadership-gouvernance, c'est-à-dire un leadership éclairé et une gouvernance de qualité, prévaut sur la malédiction des ressources naturelles. En parallèle, il s'observe également que lorsque le binôme leadership-gouvernance se détériore, comme c'est le cas notamment entre 2017 et 2018 (hausse de l'inverse de la moyenne des indices EPIN), les symptômes de la malédiction des ressources naturelles semblent se revitaliser. D'où la proposition de l'artefact de la malédiction des ressources naturelles, car la véritable cause des performances médiocres n'est pas la présence abondante ou non des ressources naturelles dans un pays. C'est plutôt et fondamentalement la qualité des institutions, et plus particulièrement la qualité du binôme leadership-gouvernance[31].

**4. Robustesse**

La présente section procède à des vérifications supplémentaires afin de s'assurer de la robustesse des résultats empiriques établis dans la section précédente. D'une part, nous évaluons la corrélation entre le binôme leadership-gouvernance et l'abondance relative en ressources naturelles. Le binôme leadership-gouvernance est mesuré comme dans la figure 4. Par contre, l'abondance relative en ressources naturelles est à présent mesurée par les exportations de carburant en pourcentage des marchandises exportées pour la période 1960-2018 (cf. figure 8). D'autre part, nous évaluons la corrélation entre l'indice de la gouvernance des ressources naturelles (2017) et l'abondance relative en ressources naturelles. L'indice gouvernance des ressources naturelles mesure la qualité de la gouvernance dans les secteurs pétrolier, gazier et minier de 81 pays. Cet indice est produit par *The Natural Resource Governance Institute* (NRGI, Classement : 0=faible et 100=élevé). L'abondance relative en ressources naturelles est mesurée d'abord (i) par les exportations de minerais et de métaux comme dans la figure 4 (cf. figure 9), puis (ii) par les exportations de carburant en pourcentage des marchandises exportées pour la période 1960-2018 (cf. figure 10).

---

[30] Émergence économique, c'est-à-dire un moment particulier de la dynamique économique correspondant à l'affirmation d'une séquence de croissance significative susceptible de déboucher sur une croissance soutenue et durable (cf. Rougier et Piveteau 2012, p. 72).

[31] Dans la littérature récente, certains chercheurs réfutent l'idée d'après laquelle le binôme leadership-gouvernance permet de mieux comprendre les performances économiques en RDC durant la période 2012-2016, voir par exemple Englebert (2016) et surtout Marysse (2018) qui propose une lecture critique de la dynamique économique en RDC durant la période 2010-2015. Dans sa conclusion, Marysse (2018, p. 253) estime que c'est davantage la qualité de l'investissement qui explique la croissance économique en RDC entre 2010-2015. Toutefois, la thèse défendue par l'auteur soulève plusieurs autres questions d'ordre analytique, dont les plus importantes sont sans doute : (i) serait-il possible d'avoir, dans un pays comme la RDC, un flux d'investissements de qualité sans un leadership éclairé et une gouvernance de qualité ? (ii) la survenance des éléphants blancs n'est-elle pas le fait de la détérioration du binôme leadership-gouvernance ? (iii) Que dire des propos suivants de North et Thomas (1973, p. 2), voir aussi Acemoglu et al. (2005, p. 388), d'après lesquels l'explication fondamentale des performances économiques réside dans les différences d'institutions ?

Figure 8 : Produit cartésien du binôme leadership-gouvernance par l'abondance relative en ressources naturelles approximées par les exportations de carburant

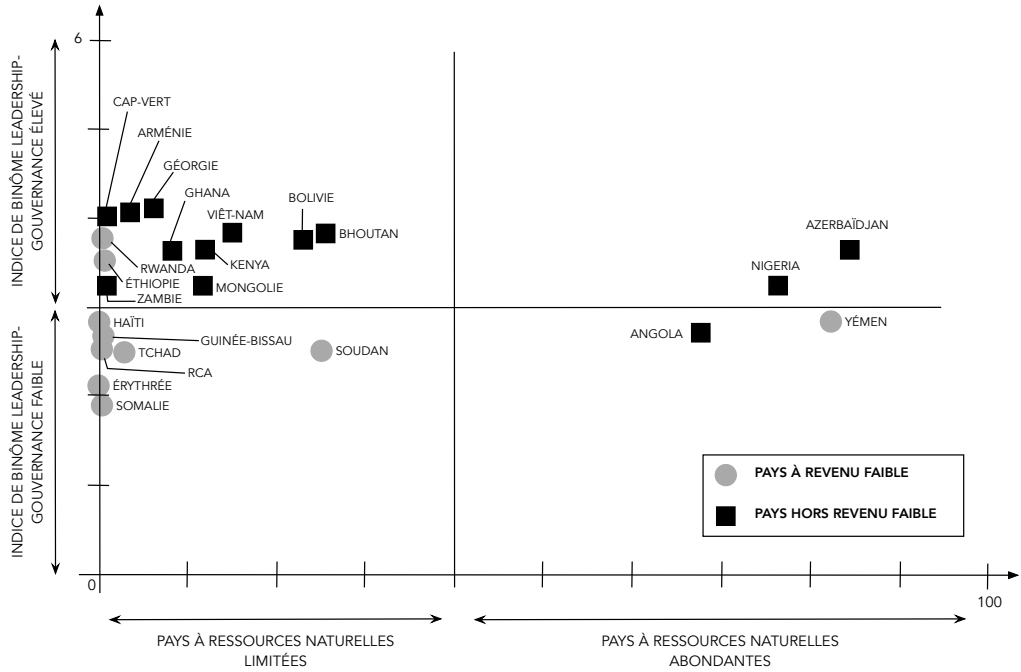

Note : Le binôme leadership-gouvernance est mesuré comme dans la figure 4. L'abondance relative en ressources naturelles est mesurée par les exportations de carburant.

Figure 9 : Produit cartésien de la gouvernance des ressources naturelles par l'abondance relative en ressources naturelles

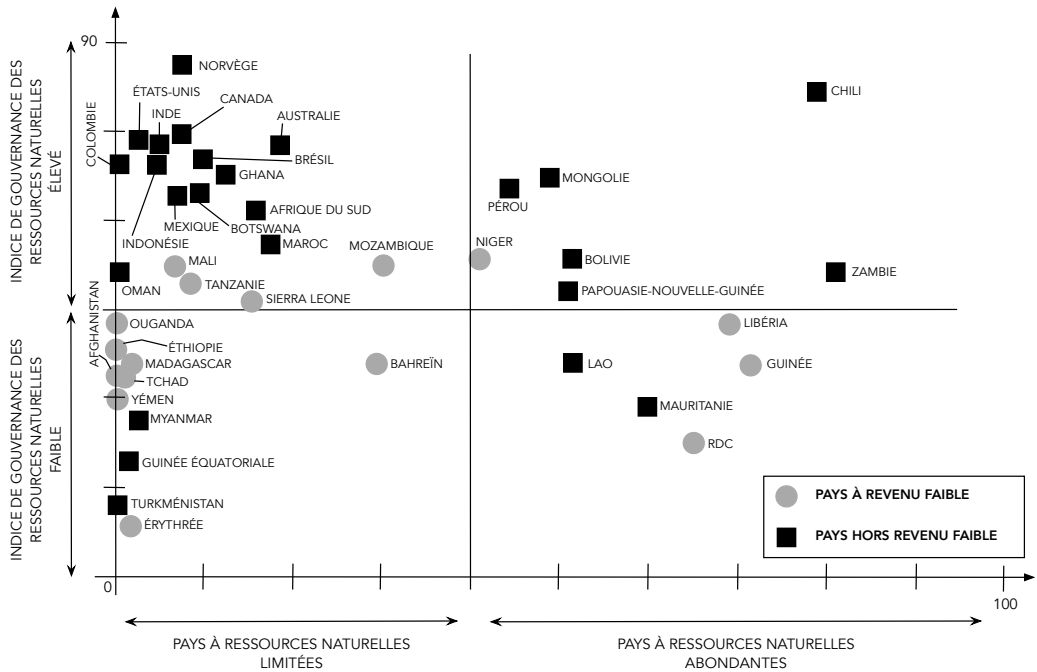

Note : Ce graphique mesure la corrélation entre l'indice de la gouvernance des ressources naturelles et l'abondance relative en ressources naturelles mesurées par les exportations de minerais et de métaux.

Figure 10 : Produit cartésien de la gouvernance des ressources naturelles par l'abondance relative en ressources naturelles approximées par les exportations de carburant

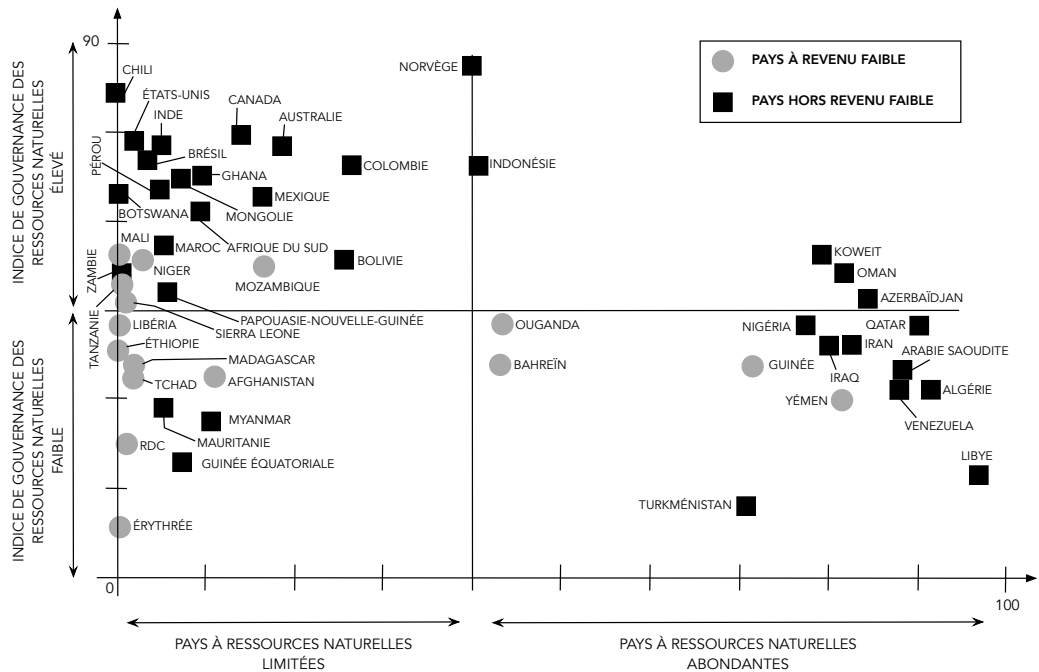

Note : Ce graphique mesure la corrélation entre l'indice de la gouvernance des ressources naturelles (2017) et l'abondance relative en ressources naturelles mesurées par les exportations de carburant (Période: 1960-2018).

Deux constants majeurs, et consistants avec les résultats établis dans la section précédente, se dégagent des figures 8, 9 et 10 :

Constant 1. Le binôme leadership-gouvernance, d'une part, aide à mieux comprendre les performances économiques réalisées par les différents pays de l'échantillon considéré et, d'autre part et plus spécifiquement, permet de lever l'ambiguïté que pose la validité empirique de l'hypothèse de la malédiction des ressources naturelles. En effet, comme il ressort clairement de la figure 8, l'abondance ou la rareté relative des ressources naturelles semble ne pas être un facteur explicatif fondamentalement pertinent dans la compréhension des performances économiques des pays. Ce résultat corrobore les évidences établies, précédemment, dans la figure 4.

Constant 2. Comme il ressort de la figure 9 et de la figure 10, l'indice de la gouvernance des ressources naturelles n'est pas un candidat susceptible d'aider à mieux cerner l'ambiguïté que pose la malédiction des ressources naturelles (cf. Note de bas de page 21, voir aussi la figure 4). Par exemple, on retrouve des pays hors revenu faible, à la fois, avec des valeurs faibles et élevées de l'indice de la gouvernance des ressources naturelles.

## 5. Conclusion

Ce papier a proposé un réexamen de la validité empirique de l'hypothèse de la malédiction des ressources naturelles, en utilisant la base de données des pays exportateurs de minerais et de métal. Après une investigation sur données, trois principales évidences ont été établies. Tout d'abord, il ressort que tous les pays riches en ressources naturelles (PRRN) n'affichent pas forcément des performances économiques, politiques et sociales médiocres. Deuxièmement, il a été noté que les PRRN avec des performances économiques, politiques et sociales médiocres possèdent tous un portefeuille d'exportations faiblement diversifié. En revanche, et en dernier lieu, il a été constaté que les PRRN avec un portefeuille d'exportations faiblement diversifié n'affichent pas forcément des performances économiques, politiques et sociales médiocres.

Ces observations nous ont donc conduit à adresser deux questions principales. La première est celle de comprendre : pourquoi certains pays riches en ressources naturelles tirent profit de leurs dotations et d'autres non ? La deuxième question que nous avons examinée est celle de préciser : quels sont les facteurs qui sous-tendent la validité de l'hypothèse de la malédiction des ressources naturelles dans certains pays en développement ? En effet soutenons-nous, en réponse à la première question, que si certains pays riches en ressources naturelles tirent profit de leurs dotations et d'autres non, c'est parce que dans la première catégorie des pays, les dirigeant font preuve d'un leadership éclairé et d'une gouvernance de qualité. Par construction analytique, un leadership et une gouvernance sont dits éclairé et de qualité respectivement si et seulement si : leadership et gouvernance se traduisent, à leur tour, par l'instauration effective, la promotion et le maintien des institutions pro-développement. À la deuxième question, nous avons noté qu'il existe plusieurs facteurs susceptibles d'expliquer la validité artefactuelle de la malédiction des ressources naturelles dans certains pays en développement. Dans ce papier, nous défendons l'idée d'après laquelle que la malédiction des ressources naturelles est artefactuellement valide dans les pays qui sont à la fois riches en ressources naturelles et dépourvus des dirigeants ayant la volonté de mettre en œuvre des réformes courageuses visant l'amélioration du bien-être de la population et la capabilité nécessaire de négocier ou de résister aux lobbyistes et groupes de pression. Autrement dit, nous soutenons qu'il n'existe pas de malédiction des ressources naturelles. Il n'existe que de défaillance du binôme leadership-gouvernance dans la gestion des ressources naturelles.

À l'effet de formaliser cet argumentaire, nous avons développé un modèle d'interactions stratégiques inspiré des jeux bayésiens. Une caractéristique essentielle du modèle que nous avons considéré est qu'il tient explicitement compte du fait que dans un pays riche en ressources naturelles, le décideur politique fait face à des groupes de pression puissants (« Léviathan »). En outre, nous avons supposé que le décideur politique est sujet à deux contraintes majeures, à noter : la contrainte de gouvernance (*lemme 1*) et la contrainte de leadership (*lemme 2*). Partant de ces considérations empiriques et analytiques, nous avons montré qu'une amélioration du leadership peut aider à discipliner le comportement du Léviathan (*théorème 1*) et, par ailleurs, à générer une amélioration de Pareto (*théorème 2*) dans la gestion des ressources naturelles. Ensuite, nous nous sommes servi des évidences tirées des données EPIN du groupe de la Banque mondiale pour étayer ces résultats théoriques. En parallèle, la base de données des pays exportateurs de carburant a été utilisée pour s'assurer de la robustesse des résultats de nos analyses.

De manière générique, les résultats de nos analyses s'inscrivent dans la lignée de ceux obtenus notamment par Robinson et al. (2006), Kabuya Kalala et Tshiunza Mbiye (2009), et Menaldo (2016) qui montrent que la prise en compte des institutions, dans l'analyse, permet de lever l'ambiguïté que pose la validité empirique de l'hypothèse de la malédiction des ressources naturelles. Toutefois, contrairement à la littérature précédemment évoquée, nous insistons sur un type particulier d'institutions, à noter le binôme leadership-gouvernance. De manière spécifique, et comme cela est en partie documenté et soutenu notamment par Dietsche (2007), Stevens et Dietsche (2008) ou encore Parker et Vadheim (2017), nous avons exploité dans notre modèle le fait que, dans la gestion des ressources naturelles, le décideur politique est confronté à des lobbyistes ou groupes d'intérêt puissants. Et de manière générale, les intérêts de ceux-ci ne sont pas toujours compatibles avec les politiques visant à promouvoir le bien-être social. Ainsi, nous avons formellement montré que pour résister aux pressions diverses provenant de ces différents lobbyistes, le décideur devra davantage faire preuve d'un leadership éclairé et d'une gouvernance de qualité. Les deux à la fois, de sorte que l'un ne doit pas aller sans l'autre et vice versa, d'où le binôme leadership-gouvernance.

En sus, nous avons prolongé l'analyse de Matata (2016) sur le binôme leadership-gouvernance, en discutant du rôle effectif de ce mécanisme dans la gestion des ressources naturelles dans les pays en développement, notamment dans la capacité pour un décideur politique à anticiper les tendances futures des cours des matières premières et à insérer les exploitations locales des ressources naturelles dans la chaîne de valeur mondiale, mais aussi dans sa capacité à tirer parti des leçons apprises ailleurs ou du passé. De cet exercice, il ressort que le succès ou l'échec de tout pays dans la gestion de ses ressources naturelles, et particulièrement de ses ressources minières, devrait être mesuré par sa capacité : (i) à se prémunir des effets induits par les chocs sur les cours des matières premières; (ii) à bâtir une économie diversifiée et résiliente; (iii) à tirer effectivement parti des leçons apprises ailleurs ou du passé; (iv) à réviser ou à changer chaque fois les stratégies de gestion si cela s'avérait nécessaire. Sur base des prédictions théoriques et des évidences empiriques, nous avons conjecturé que le binôme leadership-gouvernance permet garantir une telle capacité (cf. *Conjecture 1*).

Nous reconnaissons que le cadre d'analyse proposé dans le présent papier peut être élargie ou prolongé sur plusieurs directions de recherche. Une des directions que nous comptons emprunter dans le futur est notamment celle de généraliser les deux théorèmes du binômes leadership-gouvernance dans un environnement avec un nombre fini de dossiers, supérieur ou égal à deux, à traiter par les décideurs politiques. Une autre direction intéressante serait celle d'endogéiniser les intérêts des différents groupes de pressions ou lobbyistes qui agissent dans le secteur minier. *In fine*, notre papier a tenté de raffiner l'hypothèse de la malédiction des institutions en montrant que les performances médiocres politiques, économiques et sociales observées dans les pays riches en ressources naturelles peuvent être expliquées par le déficit du leadership et de la gouvernance. Plus particulièrement, nous soulignons que ce déficit est provoqué par l'incapacité du décideur politique à imposer les politiques économiques saines ou les mesures des réformes transformationnelles face à la pression exercée par les lobbies qui cherchent à prioriser leurs intérêts respectifs. Tout récemment, Boucekkine et al. (2019) soutiennent que la version institutionnelle de la malédiction des ressources naturelles ne peut à elle seule expliquer l'hétérogénéité observée dans l'interaction entre la croissance économique et l'abondance relative en ressources naturelles. Pour rendre compte de ce fait, les auteurs proposent un nouveau mécanisme combinant l'accès aux ressources naturelles et l'accumulation du capital humain, en référence à la « théorie de la modernisation » mise en avant par Lipset (1959, 1960). Il s'agit donc d'une piste de recherche à explorer pour au moins deux raisons : (i) Évaluer empiriquement les interactions entre le développement et le processus de démocratisation dans un cadre d'analyse qui combine à la fois la littérature sur le changement institutionnel, la littérature sur la croissance endogène, et la littérature sur la malédiction des ressources naturelles; (ii) Mieux discriminer, théoriquement et empiriquement, le soutien mixte caractérisant la littérature économique sur l'hypothèse de la malédiction des ressources naturelles.

**Bibliographie**

Ross, Michael L., 2012, *The Oil Curse: How Petroleum Wealth Shapes the Development of Nations*, Princeton, NJ: Princeton University Press, 312 pages.

Ross, Michael L., 2015, "What Have We Learned About the Resource Curse?" *Annual Review of Political Science*, vol. 18, 239-259.

Rosser, Andrew, 2006, The Political Economy of the Resource Curse: A Literature Survey, IDS Working Papers, No. 268, Brighton, Institute of Development Studies.

Rotberg, Robert I., 2015, *On Governance: What It Is, What It Means and Its Policy Uses*, Montreal: McGill-Queen's University Press, 179 pages.

Rougier, Eric et Alain Piveteau, 2012, « Comprendre l'émergence économique: convergence sectorielle, autonomie compétitive et changement institutionnel », *Mondes en développement*, vol. 158, no. 2, 71-86.

Sachs, Jeffrey D. and Andrew M. Warner, 1999, "The Big Rush, Natural Resource Booms and Growth," *Journal of Development Economics*, vol. 59, no. 1, 43-76.

Sachs, Jeffrey D. and Andrew Warner, 1995, "Natural Resource Abundance and Economic Growth," NBER Working Paper, no. 5318. Cambridge, MA: NBER.

Sala-i-Martin, Xavier and Arvind Subramanian, 2013, "Addressing the Natural Resource Curse: An Illustration from Nigeria," *Journal of African Economies*, vol. 22, no. 4, 570-615.

Savada, Andrea Matles and William Shaw, 1992, *South Korea: A Country Study*, Washington, D.C.: Federal Research Division, Library of Congress, 408 pages.

Schneider, Hartmut, 1999, "Participatory Governance: The Missing Link for Poverty Reduction," OECD Development Centre Policy Briefs 17, 30 pages.

Schumpeter, Joseph, 1942, *Capitalism, Socialism and Democracy*, Harper & Brothers, 431 pages.

Slaibi, Ahmad A. and Steven C. Kyle, 2007, "Macroeconomic Impact of Mineral Revenues on General Market Equilibrium and Poverty Alleviation in Sub-Saharan Africa," *The Journal of Energy and Development*, vol. 32, no. 2, 283-310.

Sparks, Allister, 1996, *Tomorrow Is Another Country: The Inside Story of South Africa's Road to Change*, Chicago, IL: University of Chicago Press, 261 pages.

Stephenson, N. W. and W. H. Dunn, 1940a, *George Washington*, Volume 1, New York: Oxford University Press, 473 pages.

Stephenson, N. W. and W. H. Dunn, 1940b, *George Washington*, Volume 2, New York: Oxford University Press, 596 pages.

Stevens, Paul and Evelyn Dietsche, 2008, "Resource Curse: An Analysis of Causes, Experiences and Possible ways Forward," *Energy Policy*, vol. 36, 56-65.

The Economist, 1977, *The Dutch Disease*, pp. 82-83.

Tsabora, James, 2014, "Fighting the 'Resource Wars' in the DRC: An Exploratory Diagnosis of the Legal and Institutional Problems," *The Comparative and International Law Journal of Southern Africa*, vol. 47, no. 1, 109-128.
43

# ANNEXES

A.1. Preuve du lemme 1

A.2. Preuve du lemme 2

A.3. Description des principales données utilisées

**A.1. Preuve du lemme 1**

Tout d'abord, considérons le problème du groupe de pression $GI_i$. Soit $q_i \equiv \pi_i \xi_i(1) + (1-\pi_i)\xi_i(0)$ la probabilité que le groupe de pression $GI_i$ fasse du lobbying. Soit $\Xi_i \equiv q_{-i}\gamma_i + (1-q_{-i})$ la probabilité que l'accès soit accordé au groupe de pression $GI_i$ lorsqu'il fait le lobbying. Étant donné $\theta_i$, $GI_i$ choisit $\xi_i(\theta_i)$ tel que :

$$\xi_i(\theta_i) = \underset{\xi_i(\theta_i) \in [0,1]}{\operatorname{argmax}} \mathbb{E}v_i\{\xi_i(\theta_i)\},$$

où :

$$\mathbb{E}v_i\{\xi_i(\theta_i)\} = \xi_i(\theta_i) \times [\Xi_i \times \rho_i(1,1) + (1-\Xi_i) \times \rho_i(1,0) - f_i] \\ + [1-\xi_i(\theta_i)] \times \rho_i(0,0).$$

Partie (1). Prouvons qu'un équilibre dans lequel $\xi_i(\theta_i) = \theta_i$ pour chaque $\theta_i$ et chaque $i$ existe si et seulement si :

$$\frac{\pi_1 f_1 + \pi_2 f_2}{\pi_1 \pi_2} \geq 1.$$

Pour ce faire, $\xi_i(\theta_i) = \theta_i$ pour tout $\theta_i$ et pour tout $i$. Il s'ensuit que $B_i(\lambda_i, 0) = \rho_i(\lambda_i, 0) = \lambda_i$ pour tout $\lambda_i$, et $B_i(1,1) = \rho_i(1,1) = \theta_i$. En sus, $q_i = \pi_i$ et $\Xi_i = \pi_{-i}\gamma_i + (1-\pi_{-i})$. Ainsi, la nécessité de la condition suit, car $\xi_i(0) = 0$ requiert :

$$\frac{d\mathbb{E}v_i}{d\xi_i(0)} \leq 0 \Leftrightarrow 1 - \Xi_i - f_i \leq 0.$$

Dès lors établit-on que :

$$\begin{cases} \dfrac{d\mathbb{E}v_1}{d\xi_1(0)} \leq 0 \Leftrightarrow \gamma_1 \geq 1 - \dfrac{f_1}{\pi_2} \\ \dfrac{d\mathbb{E}v_2}{d\xi_2(0)} \leq 0 \Leftrightarrow \gamma_1 \leq \dfrac{f_2}{\pi_1} \end{cases}.$$

Par conséquent :
$$\gamma_1 \in \left[1 - \frac{f_1}{\pi_2}, \frac{f_2}{\pi_1}\right].$$

Cet intervalle est non vide si et seulement si :
$$\frac{\pi_1 f_1 + \pi_2 f_2}{\pi_1 \pi_2} \geq 1.$$

Par ailleurs, la suffisance de la condition suit directement en définissant :
$$\gamma_1 \in \left[1 - \frac{f_1}{\pi_2}, \frac{f_2}{\pi_1}\right],$$

ce qui est possible étant donné que :
$$\frac{\pi_1 f_1 + \pi_2 f_2}{\pi_1 \pi_2} \geq 1,$$

avec $X_i(\rho) = 0$ pour chaque $i$, et en notant que :
$$\frac{d\mathbb{E}v_i}{d\xi_i(1)} = 1 - f_i > 0,$$

ce qui implique $\xi_i(1) = 1$.

Partie (2). Pour prouver la deuxième partie du lemme, nous procéderons en huit étapes. Supposons que :
$$\frac{\pi_1 f_1 + \pi_2 f_2}{\pi_1 \pi_2} < 1.$$

Étape 1. Nous établissons que $B_i^A(0) < 1/2$ implique $\rho_i(0,0) = 0$ pour tout $i$. Pour le montrer, procédons par contradiction. Supposons que $B_i^A(0) \geq 1/2$ pour certains $i$. Puisque $\pi_i < 1/2$, il faut que soit $\xi_i(0) > \lambda_i(1)$, soit $\xi_i(0) = \xi_i(1) = 1$. Dans les deux cas, $\xi_i(0) > 0$, $\xi_i(0) > \xi_i(1)$, et $\pi_i < 1/2$ impliquent $B_i^A(1) < 1/2$ et, par conséquent, $\rho_i(1,0) = 0$. Il en résulte :
$$\frac{d\mathbb{E}v_i}{d\xi_i(0)} = -f_i - \rho_i(0,0) < 0,$$

ce qui contredit $\xi_i(0) > 0$.

Étape 2. Nous établissons que $q_i > 0$ pour chaque $i$. Pour le montrer, procédons par contradiction. Supposons que $\xi_i(0) = \xi_i(1) = 0$ pour certains $i$. Il vient que $B_i^A(0) = \pi_i < 1/2$ et $\rho_i(0,0) = 0$. En outre, nous avons que $\Xi_{-i} = 1$ implique $\xi_{-i}(\theta_{-i}) = \theta_{-i}$ pour chaque $\theta_{-i}$. Il en résulte que $q_{-i} = \pi_{-i}$, $\Xi_i \geq 1 - \pi_{-i}$ et $X_{-i}(\rho) = 0$. Dans le même temps, $\xi_i(1) = 0$ nécessite :

$$\frac{d\mathbb{E}v_i}{d\xi_i(1)} \leq 0 \Leftrightarrow \Xi_i + (1 - \Xi_i) \times \rho_i(1,0) \leq f_i.$$

Il y a trois cas possibles.

1. $B_i^A(1) \in (0,1)$.
   Lequel implique $X_{-i}(\rho) > 0$. Puisque $X_{-i}(\rho) = 0$, il suit que $\Xi_i = 1$ et en conséquence :
   $$\frac{d\mathbb{E}v_i}{d\xi_i(1)} = 1 - f_i > 0.$$

2. $B_i^A(1) = 1$.
   Lequel implique $\rho_i(1,0) = 1$. Comme précédemment, on a que :
   $$\frac{d\mathbb{E}v_i}{d\xi_i(1)} = 1 - f_i > 0.$$

3. $B_i^A(1) = 0$.
   Lequel implique $\rho_i(1,0) = 0$. Puisque $\Xi_i \leq 1 - \pi_{-i}$, on a que :
   $$\frac{d\mathbb{E}v_i}{d\xi_i(1)} = 1 - \pi_i - f_i > 0,$$
   l'inégalité étant stricte car $\pi_{-i} < 1/2$ et
   $$\frac{\pi_1 f_1 + \pi_2 f_2}{\pi_1 \pi_2} < 1.$$
   Ainsi, il suit :
   $$\frac{d\mathbb{E}v_i}{d\xi_i(1)} > 0.$$

Dans les trois cas $d\mathbb{E}v_i/d\xi_i(1) > 0$ contredit le fait que $\xi_i(1) = 0$.

Étape 3. Nous établissons que $B_i^A(1) \geq 1/2$ pour tout $i$, et en conséquence $\xi_i(1) > \xi_i(0)$. Pour le montrer, procédons par contradiction. Supposons que $B_i^A(1) < 1/2$ pour certains $i$. De ce fait, il suit que, $\rho_i(1,0) = 0$, et donc :

$$\frac{d\mathbb{E}v_i}{d\xi_i(0)} = -f_i - \rho_i(0,0) < 0,$$

ce qui implique que $\xi_i(0) = 0$. Puisque nous avons déjà que $\xi_i > 0$, nous devons donc choisir $\xi_i(1) > 0$. Ensuite, il vient que $\xi_i(1) > 0$ et $\xi_i(0) = 0$ impliquent $B_i^A(1) = 1$. D'où, une contradiction. Donc $B_i^A(1) \geq 1/2$ pour chaque $i$. Étant donné $q_i > 0$ et $\pi_i < 1/2$, il suit que $B_i^A(1) \geq 1/2$ nécessite $\xi_i(1) > \xi_i(0)$.

Étape 4. Nous établissons $[\xi_i(1) - \xi_i(0)] < 1$ pour tout $i$. Supposons par contradiction que $\xi_i(1) = 1$ et $\xi_i(0) = 0$ pour certains $i$. Il en résulte que $X_{-i}(\rho) = 0$. En outre :

$$\frac{\pi_1 f_1 + \pi_2 f_2}{\pi_1 \pi_2} < 1$$

implique $[\xi_{-i}(1) - \xi_{-i}(0)] < 1$. De ce qui précède, il se dégage deux cas sont possibles :

1. $\xi_{-i}(0) = 0$ et $\xi_{-i}(1) \in (0,1)$.
   Dans ce cas, nous avons $B_{-i}^A(1) = 1$, ce qui implique $\rho_i(1,0) = 1$. En conséquence, il suit que :

   $$\frac{d\mathbb{E}v_{-i}}{d\xi_{-i}(1)} = 1 - f_{-i} > 0$$

   ce qui contredit le fait que $\xi_{-i}(1) < 1$.

2. $\xi_{-i}(0) > 0$.
   Dans ce cas, nous avons $B_{-i}^A(1) < 1$, ce qui implique $X_{-i}(\rho) > 0$. Puisque $X_{-i}(\rho) = 0$, il suit que $\Xi_{-i} = 1$ et :

   $$\frac{d\mathbb{E}v_{-i}}{d\xi_{-i}(0)} = -f_{-i} < 0$$

   ce qui contredit le fait que $\xi_{-i}(0) > 0$.

D'où $[\xi_i(1) - \xi_i(0)] < 1$ pour tout $i$.

Étape 5. Nous établissons $B_1^A(1) > 1/2$ et $B_2^A(1) = 1/2$. Pour le montrer, considérons $B_i^A(1) = 1/2$ pour certains $i$. Par contradiction, supposons par contradiction que $B_i^A(1) > 1/2$ pour tout $i$. Il suit que $\rho_i(1,0) = 1$, ce qui implique $\xi_i(1) = 1$ et $\xi_i(0) \in (0,1)$. Ce dernier nécessite que :

$$\frac{d\mathbb{E}v_i}{d\xi_i(0)} = 0 \Leftrightarrow 1 - \Xi_i = f_i.$$

En sus, $q_i > \pi_i$. D'une part, de la condition que $1 - \Xi_i = f_i$ pour tout $i$, il vient que :

$$\Xi_1 + f_1 = \Xi_2 + f_2 \Rightarrow \gamma_1 = \frac{q_2 + (f_2 - f_1)}{q_1 + q_2}.$$

D'autre part, la condition que $1 - \Xi_2 = f_2$ implique $\gamma_1 = f_2/q_1$. De ce fait, les deux expressions de $\gamma_1$ impliquent :

$$\frac{q_1 f_1 + q_2 f_2}{q_1 q_2} = 1,$$

ce qui est contradictoire car :

$$\frac{\pi_1 f_1 + \pi_2 f_2}{\pi_1 \pi_2} < 1$$

et $q_i = \pi_i$ pour tout $i$. Ainsi conclut-on $B_i^A(1) = 1/2$ pour certains $i$.

À présent, établissons :
$$B_2^A(1) = 1/2.$$

Pour le montrer, supposons par contradiction que $B_2^A(1) > 1/2$. De ce qui précède, il suit que $B_1^A(1) = 1/2$ implique $X_1(\rho) = 1/2 > X_2(\rho)$. Étant donné que $\alpha > 1$, nous avons $\Xi_1 = 1$, ce qui implique :

$$\frac{d\mathbb{E}v_1}{d\xi_1(0)} = -f_1 < 0.$$

Il s'ensuit que $\xi_1(0) = 0$ ce qui, ensemble avec $q_1 > 0$, implique $\xi_1(1) > 0$ et $B_1^A(1) = 1$, d'où une contradiction. En conséquence, $B_2^A(1) = 1/2$, lequel qui implique $X_2(\rho) = 1/2$. En dernier lieu, établissons que $B_1^A(1) > 1/2$ et $\rho_1(1,0) = 1$. Supposons par contradiction que $B_1^A(1) = 1/2$. Il en résulte que $X_1(\rho) = 1/2$. Puisque $X_2(\rho) = 1/2$ et $\alpha > 1$, il vient que $\Xi_1 = 1$, ce qui implique :

$$\frac{d\mathbb{E}v_1}{d\xi_1(0)} = -f_1 < 0,$$

et par conséquent $\xi_1(0) = 0$. Ce dernier, associé au fait que $q_1 > 0$, implique $\xi_1(1) > 0$ et $B_1^A(1) = 1$, d'où une contradiction. En conséquence conclut-on que $B_1^A(1) > 1/2$ implique $\rho_1(1,0) = 1$.

Étape 6. Nous établissons $\xi_i(1) = 1$ pour tout $i$. En effet, on peut le voir directement pour $GI_1$ étant donné que $\rho_1(1,0) = 1$ implique que :

$$\frac{d\mathbb{E}v_1}{d\xi_1(1)} = 1 - f_1 > 0.$$

Considérons à présent $GI_2$. Supposons par contradiction que $\xi_2(1) \in (0,1)$. En effet, étant donné que $q_2 > 0$ et $\pi_2 < 1/2$, $B_1^A(1) = 1/2$ implique $\xi_2(0) \in (0,1)$. Étant donné $\xi_2(\theta_2) \in (0,1)$ pour $\theta_2$, il faut que :

$$\begin{cases} \dfrac{d\mathbb{E}v_2}{d\xi_2(1)} = 0 \Leftrightarrow \Xi_2 + (1 - \Xi_2) \times \rho_2(1,0) = f_2 \\ \dfrac{d\mathbb{E}v_2}{d\xi_2(0)} = 0 \Leftrightarrow (1 - \Xi_2) \times \rho_2(1,0) = f_2. \end{cases}$$

En effet, remarquons que les deux égalités précédentes ne peuvent être satisfaites simultanément que si :
$$\Xi_2 = 0,$$

ce qui n'est possible que si $q_1 = 1$. D'où contradiction avec le fait que $\xi_1(1) > \xi_1(0)$.

Étape 7. Soit $\xi_i(0)$ pour tout $i$. Soit :

$$B_i^A(1) = \frac{\pi_i}{[\pi_i + (1 - \pi_i)\xi_i(0)]}.$$

Pour $GI_2$, il vient que $B_2^A(1) = 1/2$, ce qui implique :

$$\xi_2(0) = \left(\frac{\pi_2}{1 - \pi_2}\right) \in (0,1).$$

Pour $GI_1$, il vient que $B_1^A(1) > 1/2$, ce qui implique $\rho_1(1,0) = 1$. Étant donné que $[\xi_i(1) - \xi_i(0)] < 1$ (cf. *Étape 4*), $\xi_1(1) = 1$ et $\xi_1(1) > \xi_1(0)$ impliquent $\xi_1(0) \in (0,1)$. Pour $\xi_i(0) \in (0,1)$ pour tout $i$, il faut que $\gamma_1 \in (0,1)$ et, par conséquent, $X_1(\rho) \times \alpha = X_2(\rho)$. Puisque $X_1(\rho) = 1 - B_1^A(1)$ et $X_2(\rho) = 1/2$, il découle finalement que :

$$\xi_2(0) = \left(\frac{\pi_2}{1 - \pi_1}\right) \times \left(\frac{1}{2\alpha - 1}\right) \in (0,1).$$

Étape 8. En dernier lieu, il reste à déterminer $\gamma_i$ et $\rho_i(1,0)$ pour chaque $i$. En effet, la probabilité espérée que l'accès à un groupe de pression $GI_i$ soit accordé s'il fait le lobbying est donné par :

$$\Xi_i \equiv q_{-i} \times \gamma_i + (1 - q_{-i}).$$

Étant donné $\xi_{-i}(1) = 1$ et

$$\xi_{-i}(0) = \left(\frac{\pi_{-i}}{1 - \pi_{-i}}\right) \times \left(\frac{1}{2\alpha_{-i} - 1}\right)$$

où $\alpha_1 = \alpha$ et $\alpha_2 = 1$, nous avons :

$$q_{-i} = \frac{2\alpha_{-i}\pi_{-i}}{2\alpha_{-i} - 1}.$$

En outre, puisque $\xi_i(0) \in (0,1)$ pour chaque $i$, il faut que pour chaque $i$ :

$$\frac{d\mathbb{E}v_i}{d\xi_i(0)} = 0 \Leftrightarrow (1 - \Xi_i) \times \rho_i(1,0) = f_i.$$

De ce qui précèdent, il en résulte que pour tout

$$\gamma_{-i} \times \rho_i(1,0) = \frac{2\alpha_{-i} - 1}{2\alpha_{-i}\pi_{-i}} \times f_i. \tag{A-1}$$

Nous savons par l'étape 5 que $\rho_i(1,0) = 1$. Ainsi découle-t-il de l'équation (A-1) que pour $i = 1$ :

$$\gamma_2 = \frac{f_1}{2\pi_2} \in (0,1). \tag{A-2}$$

Ensuite, sachant que $\gamma_1 = 1 - \gamma_2$, l'équation (A-2), pour $i = 2$, donne :

$$\rho_2(1,0) = \frac{\pi_2 \times (2\alpha - 1) \times f_2}{\pi_1\alpha \times (2\pi_2 - f_1)} \in (0,1). \blacksquare$$

### A.2. Preuve du lemme 2

Tout d'abord, considérons le problème du groupe de pression $GI_i$. Étant donné $\theta_i$, $GI_i$ choisit $\xi_i(\theta_i)$ tel que :

$$\xi_i(\theta_i) = \underset{\xi_i(\theta_i) \in [0,1]}{\operatorname{argmax}} \mathbb{E}v_i\{\xi_i(\theta_i)\},$$

où, étant donné $\rho_i(\lambda_i, \lambda_{-i}; \psi_i, \psi_{-i})$ :

$$\mathbb{E}v_i\{\xi_i(\theta_i)\} = \xi_i(\theta_i) \times [q_{-i} \times \gamma_i \times \rho_i(1,1; 1,0) + \cdots$$
$$\ldots + (1 - \gamma_i) \times [\pi_{-i} \times \xi_{-i}(1) \times \rho_i(1,1; 0,1) + \cdots$$
$$\ldots + (1 - \pi_{-i}) \times \xi_{-i}(0) \times \rho_i(1,1; 0,1)] + \cdots$$
$$\ldots + (1 - q_{-i}) \times \rho_i(1,0; 1,0) - f_i]$$
$$+ [1 - \xi_i(\theta_i)] \times [\pi_{-i} \times \xi_{-i}(1) \times \rho_i(0,1; 0,1) + \cdots$$
$$\ldots + (1 - \pi_{-i}) \times \xi_{-i}(0) \times \rho_i(0,1; 0,1) + \cdots$$
$$\ldots + (1 - q_{-i}) \times \rho_i(0,0; 0,0)].$$

Prouvons que quand $f_2 > (1 - \pi_1)$, il existe un équilibre tel que :

$$\xi_1(1) = 1,$$

et

$$\xi_1(0) = \xi_2(1) = \xi_2(0) = 0.$$

Dans ce cas :

$$q_1 = \pi_1,$$

et

$$q_2 = 0.$$

En sus :

$$B_1^A(1) = 1,$$

$$B_1^A(0) = 0,$$

et

$$B_2^A(0) = \pi_2.$$

Ainsi pouvons-nous déduire de ces croyances que :

$$\begin{cases} \rho_1(1,1; 0,1) = 1 \\ \rho_1(0,\cdot; 0,\cdot) = \rho_2(0,\cdot; 0,\cdot) = \rho_2(1,1; 1,0) = 0 \\ \rho_1(1,\cdot; 1,0) = \theta_1 \\ \rho_2(1,0; 1,0) = \theta_2 \text{ and } \rho_2(1,1; 0,1) \leq 1 - \theta_1. \end{cases}$$

Soit $B_2^A(1)$ une valeur quelconque dans $[0,1]$ tel que $\rho_2(1,1; 0,1) \in [0, 1 - \theta_1]$ est consistant avec $B_2^A(1)$. Puisque le groupe de pression $GI_1$ ne fait le lobbying que si et seulement si $\theta_1 = 1$, il suit que $X_1(\rho) \times \alpha = X_2(\rho) = 0$. En outre, et $\gamma_i$ peut prendre n'important quelle valeur dans $[0,1]$, $\gamma_1 + \gamma_2 = 1$.

Ainsi établit-on que :

$$\begin{cases} \dfrac{\mathrm{d}\mathbb{E}v_1}{\mathrm{d}\xi_1(1)} = 1 - f_1 > 0 \\ \dfrac{\mathrm{d}\mathbb{E}v_1}{\mathrm{d}\xi_1(0)} = -f_1 < 0 \\ \dfrac{\mathrm{d}\mathbb{E}v_2}{\mathrm{d}\xi_2(1)} = (1 - \pi_1) - f_2 < 0 \\ \dfrac{\mathrm{d}\mathbb{E}v_2}{\mathrm{d}\xi_2(0)} = -f_2 < 0, \end{cases}$$

résultat qui est cohérent avec les stratégies de lobbying.

Prouvons maintenant que quand $f_2 > (1 - \pi_1)$, dans chaque équilibre nous avons :

$$\xi_i(1) = 1,$$

et

$$\xi_i(0) = \xi_{-i}(1) = \xi_{-i}(0) = 0,$$

pour tout $i$. Pour ce faire, nous procéderons en trois étapes.

Étape 1. Dans ce cas, on a : $\xi_i(1) > \xi_i(0)$ et $\xi_i(0) < 1$ pour tout $i$. Procédons par contradiction. Supposons que :

$$\xi_i(0) \in (\xi_i(1), 1],$$

ou

$$\xi_i(1) > \xi_i(0) = 1.$$

Dans les deux cas :

$$\xi_i(0) > 0$$

et

$$B_i^A(1) < 1/2.$$

Il suit que :

$$\frac{\mathrm{d}\mathbb{E}v_i}{\mathrm{d}\xi_i(0)} = -f_i < 0,$$

ce qui contredit le fait que $\xi_i(0) > 0$.

Étape 2. Établissons le fait $\xi_1(\theta_1) = \theta_1$, pour tout $\theta_1$, implique $\xi_2(\theta_2) = 0$, pour tout $\theta_2$. Étant donné $\xi_1(\theta_1) = \theta_1$, pour tout $\theta_1$, il suit que $q_1 = \pi_1$. En outre, étant donné $\alpha > 1$, $B_1^A(1) = 1$ et $B_1^A(0) = 0$ implique $\rho_1(1,1;0,1) = 1$. Il en résulte que :

$$\frac{d\mathbb{E}v_2}{d\xi_2(0)} \geq \frac{d\mathbb{E}v_2}{d\xi_2(1)} = (1 - \pi_1) - f_2 - (1 - \pi_1) \times \rho_2(0,0;0,0) < 0,$$

ce qui implique $\xi_2(\theta_2) = 0$, pour tout $\theta_2$.

Étape 3. Établissons le fait que $[\xi_1(1) - \xi_1(0)] < 1$ implique $\xi_1(\theta_1) = 0$, pour tout $\theta_1$, et $\xi_2(\theta_2) = \theta_2$, pour tout $\theta_2$. Ici, il y a trois cas à considérer :

1. $\xi_1(1) = 1$.
   Dans ce cas, $\xi_1(0) \in (0,1)$ et $q_1 < 1$. Il en résulte que $B_1^A(1) < 1$. En outre, $B_1^A(0) = 0$, ce qui implique $\rho_1(0,\cdot;0,\cdot) = 0$ et $\rho_2(1,1;0,1) = \theta_2$. Nous obtenons alors :

$$\begin{cases} \dfrac{d\mathbb{E}v_2}{d\xi_2(1)} = q_1\gamma_2 \times \rho_2(1,1;1,0) + (1 - \gamma_2) \times (1 - \pi_1) \times \xi_1(0) \times \ldots \\ \qquad\qquad \ldots \times \rho_2(1,1;0,1) + (1 - q_1) - f_2 \\ \dfrac{d\mathbb{E}v_2}{d\xi_2(0)} = (1 - \gamma_2) \times (1 - \pi_1) \times \xi_1(0) \times \rho_2(1,1;0,1) - f_2. \end{cases}$$

Étant donné que $f_2 > (1 - \pi_1)$, nous avons :

$$\frac{d\mathbb{E}v_2}{d\xi_2(0)} < 0$$

et donc $\xi_2(0) = 0$. Ensuite, il faut que $\xi_2(1) > 0$, sinon :

$$\frac{d\mathbb{E}v_1}{d\xi_1(0)} = -f_1 < 0,$$

ce qui contredirait le fait $\xi_1(0) > 0$. Puisque $\xi_2(1) > 0 = \xi_2(0)$ implique que $B_2^A(1) = 1$ et, conjointement avec $B_1^A(1) < 1$, que $X_2(\rho) < X_1(\rho) \times \alpha$. Dès lors, $\gamma = 1$ et $d\mathbb{E}v_1/d\xi_1(0) = -f_1 < 0$, contredisent le fait que $\xi_1(0) > 0$.

2. $\xi_1(1) \in (0,1)$.
   Nous remarquons d'abord que $\xi_1(0) = 0$. Nous savons que $\xi_1(0) < 1$. Supposons par contradiction que $\xi_1(0) \in (0,1)$. Alors, il faut que :

$$\frac{d\mathbb{E}v_1}{d\xi_1(1)} = \frac{d\mathbb{E}v_1}{d\xi_1(0)} = 0,$$

ou que :
$$q_2 \gamma_2 = 1.$$

Ce dernier nécessite $q_2 = 1$. Or nous savons que cela ne peut être le cas. D'où :
$$\xi_1(0) = 0.$$

Soit $\xi_1(1) > \xi_1(0) = 0$ et $\pi_1 < 1/2$, il suit que $B_1^A(1) = 1$ et $B_1^A(0) < 1/2$ et, par conséquent, $\rho_1(1,1; 0,1) = 1$ et $\rho_1(0,\cdot; 0,\cdot) = 0$. Il en résulte que :

$$\frac{d\mathbb{E}v_1}{d\xi_1(1)} = 1 - f_1 > 0,$$

ce qui contredit le fait que $\xi_1(1) < 1$.

3. $\xi_1(1) = 0$.
   Puisque $\xi_1(1) \geq \xi_1(0)$, on a alors $\xi_1(0) = 0$. Il suit que $B_1^A(1) = \pi_1 < 1/2$, et donc $\rho_i(0,\cdot; 0,\cdot) = 0$. En sus :

$$\frac{d\mathbb{E}v_2}{d\xi_2(0)} \leq -f_2 < 0,$$

ce qui implique $\xi_2(0) = 0$. Il en résulte que $B_1^A(0) < 1/2$ et, par conséquent $\rho_i(0,\cdot; 0,\cdot) = 0$. Ainsi découle-t-il que :

$$\frac{d\mathbb{E}v_2}{d\xi_2(1)} = 1 - f_2 > 0,$$

ce qui implique $\xi_2(1) = 1$.

Prouvons à présent que quand $f_2 > (1 - \pi_1)$, il existe un équilibre tel que :
$$\xi_i(\theta_i) = \theta_i,$$

pour tout $\theta_i$ et pour tout $i$. Dans ce cas, $q_i = \pi_i$. En sus, $B_i^A(\lambda_i) = \lambda_i$ pour tout $\lambda_i$ et pour tout $i$, tel que :

1. $X_1(\rho) \times \alpha = X_2(\rho)$, lequel implique :
$$\gamma_1 = (1 - \gamma_2) \in \left(1 - \frac{f_1}{\pi_2}, 1\right];$$

2. $\rho_i(1,1; 0,1) = 1$ si $\theta_{-i} = 0$;

3. $\rho_i(0,\cdot;0,\cdot) = 0$;

4. $\rho_1(1,1;1,0) = \theta_1$ et $\rho_1(1,1;0,1) = 1$.

Il en résulte que :

$$\frac{d\mathbb{E}v_i}{d\xi_i(1)} \geq 0$$

et

$$\frac{d\mathbb{E}v_i}{d\xi_i(0)} < 0$$

pour tout $i$, ce qui est consistant avec le fait que :

$$\xi_i(\theta_i) = \theta_i,$$

pour tout $\theta_i$ et pout tout $i$. La construction de l'équilibre s'achève en laissant les stratégies d'accès et de choix de politique restantes telles que spécifiées dans le texte (cf. *Section 3*).

Finalement, prouvons que quand $f_i < 1 - \pi_i$ pour tout $i$, dans chaque équilibre nous avons :

$$\xi_i(\theta_i) = \theta_i$$

pour chaque $\theta_i$ et chaque $i$. Pour ce faire, nous procéderons en sept étapes.

Étape 1. Montrons que $B_i^A(0) = 1/2$, pour tout $i$. Procédons par contradiction. Supposons que $B_i^A(0) \geq 1/2$ pour certains $i$. Ceci est possible seulement si soit $\xi_i(1) < \xi_i(0)$, soit $\xi_i(1) = \xi_i(0) = 1$. Dans les deux cas, $B_i^A(1) < 1/2$ et $d\mathbb{E}v_i/d\xi_i(0) \leq -f_i < 0$, ce qui contredit $\xi_i(0) > 0$.

Étape 2. Montrons que $\xi_i(1) \geq \xi_i(0)$ pour chaque $i$. Supposons par contradiction que $\xi_i(1) \geq \xi_i(0)$ pour certains $i$. Cela implique $B_i^A(1) < 1/2$, ce qui est contradictoire en suivant le même argument que dans la première étape.

Étape 3. Montrons que $\xi_i(0) < 1$ pour chaque $i$. Supposons par contradiction que $\xi_i(0) = 1$ pour certains $i$. Puisque nous savons déjà que $\xi_i(1) \geq \xi_i(0)$ (cf. *Étape 2*), nous avons $\xi_i(1) = \xi_i(0) = 1$ et, par conséquent, $B_i^A(1) = \pi_i < 1/2$, ce qui est, une fois de plus, contradictoire en suivant le même argument que dans la première étape.

Étape 4. Montrons que $\xi_i(1) > 0$ pour chaque $i$. Supposons par contradiction que $\xi_i(1) = 0$ pour certains $i$. Puisque nous savons déjà que $\xi_i(1) \geq \xi_i(0)$ (cf. *Étape 2*), nous avons $\xi_i(1) = \xi_i(0) = 0$ et, par conséquent, $q_i = 0$. Puisque $B_i^A(0) = \pi_i < 1/2$, il suit que $\xi_{-i}(\theta_{-i}) = \theta_{-i}$ pour chaque $\theta_{-i}$. Et donc $d\mathbb{E}v_i/d\xi_i(1) \geq (1-\pi_i) - f_i > 0$, ce qui contredit le fait que $\xi_i(1) = 0$.

Étape 5. Montrons que $\xi_i(1) > \xi_i(0)$ pour chaque $i$. Nous savons que $\xi_i(1) \geq \xi_i(0)$ pour chaque $i$ (cf. *Étape 2*). Supposons par contradiction que $\xi_i(1) = \xi_i(0)$ pour certains $i$. Puisque nous savons aussi que $\xi_i(1) \geq 0$ (cf. *Étape 4*), il vient que $B_i^A(1) = \pi_i < 1/2$, ce qui est, une fois de plus, contradictoire en suivant le même argument que dans la première étape. En outre, il en résulte que $q_i \in (0,1)$ pour chaque $i$.

Étape 6. Montrons que $\xi_i(1) = 1$ pour chaque $i$. Supposons par contradiction que $\xi_i(1) < 1$ pour certains $i$. Nous savons que $\xi_i(1) > \xi_i(0)$ pour chaque $i$ (cf. *Étape 5*). Il suit que $\xi_i(1) \in (0,1)$, et en conséquence $d\mathbb{E}v_i/d\xi_i(1) = 0$ et $B_i^A(1) = 1$. Si $i = 1$, $B_1^A(1) = 1$ et $\alpha > 1$ impliquent $d\mathbb{E}v_1/d\xi_1(1) = 1 - f_1 > 0$, ce qui contredit le fait que $\xi_i(1) < 1$. Si $i = 2$, $d\mathbb{E}v_2/d\xi_2(1) = 0$ tient seulement si $(1 - q_1) \leq f_2$. Cette inégalité, prise ensemble avec $f_2 < 1 - \pi_1$, nécessite que $\xi_1(0) > 0$. À son tour, $\xi_1(0) > 0$ nécessite que $\gamma_1 < 1$ et en conséquence $X_1(\rho) \times \alpha < X_2(\rho)$. Ensuite, une manipulation algébrique simple montre que $B_2^A(1) = 1$ et $\xi_1(0) > 0$ impliquent $X_1(\rho) \times \alpha > X_2(\rho)$, ce qui est contradictoire.

Étape 7. Finalement, montrons que $\xi_i(0) = 0$ pour chaque $i$. Supposons par contradiction que $\xi_i(0) > 0$ pour certains $i$. Étant donné que $\xi_i(1) > \xi_i(0)$ pour chaque $i$ (cf. *Étape 5*), il suit que $\xi_i(0) \in (0,1)$. En outre, étant donné que $\xi_{-i}(1) = 1$, $\xi_i(0) > 0$ nécessite également que $\xi_{-i}(0) \in (0,1)$. Ainsi, pour chaque $k$, $\xi_k(0) \in (0,1)$ requiert que $\gamma_k(0) \in (0,1)$, ce qui nécessite que $X_1(\rho) \times \alpha = X_2(\rho)$. Enfin, étant donné $\alpha > 1$, une manipulation algébrique simple permet d'établir que $X_1(\rho) > 0$ et $X_1(\rho) \geq X_2(\rho)$, ce qui contredit le fait que $X_1(\rho) \times \alpha = X_2(\rho)$. ∎

### A.3. Description des principales données utilisées

1  **PIB par habitant**
   PIB par habitant ($ US constants de 2010)     NY.GDP.PCAP.KD
   Le PIB par habitant est le produit intérieur brut divisé par la population en milieu d'année. Le PIB est la somme de la valeur ajoutée brute de tous les producteurs résidents d'une économie plus toutes taxes sur les produits et moins les subventions non incluses dans la valeur des produits. Elle est calculée sans effectuer de déductions pour la dépréciation des biens fabriqués ou la perte de valeur ou la dégradation des ressources naturelles. Les données sont en dollars américains constants de 2010.
   Source : Données des comptes nationaux de la Banque mondiale et fichiers de données des comptes nationaux de l'OCDE, WDI 2019.
   URL: http://donnees.banquemondiale.org/indicateur/NY.GDP.PCAP.KD

2  **Ressources naturelles: Minerais et de métaux**
   Exportations de minerais et de métaux (% des     TX.VAL.MMTL.ZS.UN
   marchandises exportées)
   Les minerais et les métaux comprennent les produits des sections 27 de la CTCI (engrais bruts, minéraux); 28 (minerais métalliques, matériaux de récupération); et 68 (métaux non ferreux).

Source : Estimations du personnel de la Banque mondiale issues de la base de données Comtrade maintenue par la Division des Statistiques des Nations Unies, WDI 2019. Comtrade ou UN COMTRADE est le pseudonyme de la base de données des Nations Unies sur les statistiques du commerce international.
URL: donnees.banquemondiale.org/indicateur/TX.VAL.MMTL.ZS.UN

3 **Ressources naturelles: Carburant**
Exportations de carburant (% des marchandises exportées)     NY.GDP.PCAP.KD

Les combustibles comprennent les produits de la section 3 de la classification type pour le commerce international (CTCI: Combustibles minéraux, lubrifiants et matériaux connexes).
Source : Estimations du personnel de la Banque mondiale issues de la base de données Comtrade maintenue par la Division des Statistiques des Nations Unies, WDI 2019.
URL: http://donnees.banquemondiale.org/indicateur/NY.GDP.PCAP.KD

4 **Gouvernance des ressources naturelles**
Indice de gouvernance des ressources naturelles     RGI

L'Indice de gouvernance des ressources naturelles évalue les politiques et les pratiques auxquelles les autorités ont recours pour régir leurs industries pétrolières, gazière et minière. À chaque évaluation correspond une note composite. L'indice évalue le secteur pétrolier et gazier et le secteur minier.
Source: The Natural Resource Governance Institute (NRGI).
URL: http://resourcegovernance.org

5 **Binôme du leadership-gouvernance**
Moyenne arithmétique des scores des indices EPIN     Calculé par les auteurs du Groupe de la banque mondiale.

6 **Troïka stratégique**
Réunion de Troïka stratégique     Troïka

En dépit d'être véritablement une innovation institutionnelle dans la coordination de l'action gouvernementale, la Troïka stratégique constitue une plateforme d'interactions avec trois fonctions principales : (i) renforcer la cohésion de l'exécutif dans la mise en œuvre du programme gouvernemental; (ii) assurer une certaine homogénéisation de l'information et par conséquent, une amélioration de la prise de décisions consensuelles; (iii) assouplir le travail du conseil des ministres. Par exemple, entre 2012 et 2016, 252 réunions de Troïka stratégique ont été organisées, soit une moyenne mensuelle de 4,20 réunions. Alors qu'il n'y a eu que 60 réunions du Conseil des Ministres, soit une moyenne mensuelle d'une réunion au cours de la même période.
Source: Cabinet du Premier ministre de la RDC